%% file: Dust_I.tex
\documentclass[a4paper,usenatbib,times]{mn2e}
\usepackage{natbib,graphicx,amsmath,amsfonts,amssymb,times,txfonts}

\def\vg{\textbf{v}_{\mathrm{g}}}
\def\vd{\textbf{v}_{\mathrm{d}}}
\def\rhog{\rho_{\mathrm{g}}}
\def\hrhog{\hat{\rho}_{\mathrm{g}}}
\def\rhod{\rho_{\mathrm{d}}}
\def\hrhod{\hat{\rho}_{\mathrm{d}}}

\def\ra{\textbf{r}_{a}}
\def\va{\textbf{v}_{a}}

\def\vb{\textbf{v}_{b}}

\def\vk{\textbf{v}_{k}}

\def\Pga{P_a}
\def\rhoga{\rho_a}
\def\hrhoga{\hat{\rho}_{a}}

\def\hrhoda{\hat{\rho}_{i}}
\def\thetaa{\theta_{a}}
\def\ma{m_{a}}

\def\doha{\left(h_{a}  \right)}

\def\omga{\Omega_{a} }

\def\rhogb{\rho_{b}}
\def\sgb{s_{b}}
\def\ugb{u_{b}}
\def\Pgb{P_{b}}
\def\hb{h_{b}}
\def\hrhogb{\hat{\rho}_{b}}

\def\hrhodb{\hat{\rho}_{{\mathrm{d},b}} }
\def\thetab{\theta_{b}}
\def\mb{m_{b}}

\def\dohb{\left(h_{b}  \right)}

\def\omgb{\Omega_{b} }
\def\omdb{\Omega_{\mathrm{d},b} }

\def\mj{m_{j}}

\def\mk{m_{k}}

\def\mc{m_{c}}

\def\fp{\textbf{F}^{\rm V}_{\mathrm{drag}}}

\def\cs{c_{\mathrm{s}}}

\def\md{m_{\mathrm{d}}}

\def\kak{K_{ak}}
\def\vak{\textbf{v}_{ak}}
\def\hrak{\hat{\textbf{r}}_{ak}}

\def\hrbi{\hat{\textbf{r}}_{bi}}

\def\hraj{\hat{\textbf{r}}_{aj}}

\def\Pg{P_{\mathrm{g}}}

\def\dst{\displaystyle}

\newcommand{\kn}[2]{W_{\!\!\ #1\!\! \ #2}}
\newcommand{\gkn}[3]{\nabla_{\!\! \ #1}W_{\!\! \ #2\!\! \ #3}}

\title[Dusty gas with SPH --- I.]{Dusty gas with SPH --- I. Algorithm and test suite}

\author[Laibe \& Price]{Guillaume Laibe, Daniel J. Price \\
Monash Centre for Astrophysics (MoCA) and School of Mathematical Sciences, Monash University, Clayton, Vic 3800, Australia
}
\pagerange{\pageref{firstpage}--\pageref{lastpage}} \pubyear{2011}

\begin{document}
\include{journaux}

\label{firstpage}
\bibliographystyle{mn2e}
\maketitle

\begin{abstract}
We present a new algorithm for simulating two-fluid gas and dust mixtures in Smoothed Particle Hydrodynamics (SPH), systematically addressing a number of key issues including the generalised SPH density estimate in multi-fluid systems, the consistent treatment of variable smoothing length terms, finite particle size, time step stability, thermal coupling terms and the choice of kernel and smoothing length used in the drag operator. We find that using double-hump shaped kernels improves the accuracy of the drag interpolation by a factor of several hundred compared to the use of standard SPH bell-shaped kernels, at no additional computational expense. In order to benchmark our algorithm, we have developed a comprehensive suite of standardised, simple test problems for gas and dust mixtures: \textsc{dustybox}, \textsc{dustywave}, \textsc{dustyshock}, \textsc{dustysedov} and \textsc{dustydisc}, the first three of which have known analytic solutions.

We present the validation of our algorithm against all of these tests. In doing so, we show that the spatial resolution criterion $\Delta \lesssim c_{\rm s} t_{\rm s}$ is a necessary condition in all gas+dust codes that becomes critical at high drag (i.e. small stopping time $t_{\rm s}$) in order to correctly predict the dynamics. Implicit timestepping and the implementation of realistic astrophysical drag regimes are addressed in a companion paper.
\end{abstract}


\section{Introduction}
\label{sec:intro}
 Most of our observational information regarding the interstellar medium comes to us via dust. Over the last few years, observations using the \emph{Spitzer} and \emph{Herschel} space telescopes have substantially improved our observational sensitivity to and resolution of dust emission in a wide range of astrophysical environments. Dust grains provide the materials from which the solid cores required for the planet formation process are built (see e.g. \citealt{ChiangYoudin2010}). They also modify the dynamical evolution of the surrounding gas by exchanging momentum and energy via microscopic collisions \citep{Epstein1924,Baines1965}. Dust grains are also the main sources of  the opacities in star-forming molecular clouds, thus determining their evolution by controlling the thermodynamics. Accurate determination of both the dynamics of the star and planet formation process and its observational signature thus require modelling the coupled evolution of gas and dust.
 
 Given that the N-body evolution of solid particles in a mixture of gas and solid material would be prohibitive in terms of both physical complexity and computational cost, the usual approach is to regard the solid phase as a continuum and the mixture as a two-fluid system coupled by a drag term. This requires averaging physical quantities over a control volume $V$ that is large enough to be statistically meaningful but sufficiently small compared to the macroscopic scale to allow a continuum description. In diluted astrophysical media, the frequency of collisions between dust particles are infrequent enough that the intrinsic pressure of the dust phase can be regarded as negligible to a very good level of approximation, leaving the dust as a free-streaming collisionless fluid whose motion is controlled solely by gravitational forces and the drag-term interaction with the gas.

 However, even with a continuous description of the mixture, the equations can be solved analytically only for a few simple cases (the solutions to two specific problems, \textsc{dustybox} and \textsc{dustywave}, corresponding to mutually interpenetrating fluids and acoustic waves propagating in a dusty gas, respectively, are derived in \citealt{LP11a}). As a result, numerical codes have been developed in order to model more realistic systems based either on $N$-body dust particles in Eulerian grid-based hydrodynamics (e.g.  \citealt{Fromang2006,PM2006,Johansen2007,Miniati2010,Bai2010}) or with a two-fluid Smoothed Particle Hydrodynamics (SPH) approach.
 
  SPH methods for simulating two-fluid mixtures were first developed by \citet{Monaghan1995}, improved (via an implicit treatment of the drag terms) in \citet{Monaghan1997} and applied in an astrophysical context to the dynamics of dust grains in protoplanetary disks \citep{Maddison2003,Rice04,BF2005}. The particle-based nature of the SPH formalism means that the addition of a dusty fluid is natural. More importantly, the drag term that couples the two phases can be implemented such that the total linear and angular momentum of the system are exactly (and simultaneously) conserved, in line with the Hamiltonian and exactly conservative nature of the core SPH method \citep[e.g.][]{Price2011}.
  
   However, the standard methods for treating dusty gas in SPH were developed over 15 years ago and our initial attempts to simply apply the existing formulations uncovered several issues that needed to be addressed. Specifically: 1) the original formulations assumed a spatially constant SPH smoothing length; 2) the SPH terms for the conservative part of the equations should be derived from a Lagrangian;  3) we found that the use of the standard cubic spline kernel for drag terms could be significantly inaccurate; 4) we encountered several previously unexplored resolution issues in simulating two-fluid mixtures; 5) aspects of the implicit timestepping scheme suggested by \citet{Monaghan1997} were found to be problematic; 6) that treatments of drag have to date generally limited to linear drag regimes; and finally 7) that the existing schemes --- having been developed with both astrophysical and geophysical dust problems in mind --- have not been widely benchmarked on problems appropriate to astrophysics; Indeed there is a general lack of standardised test problems for two-fluid dust/gas codes, a problem partially addressed by our first paper \citep{LP11a}.
   
In this and a companion paper (\citealt{LP11c}, hereafter Paper~II), we set out to systematically address issues 1)--7) in order to develop a robust and accurate code for simulating the dynamics of dust in star and planet formation. The importance of modelling the dust-gas interaction has been highlighted by recent studies showing that instabilities in dust-gas mixtures are good candidates for triggering the concentration of dust during planetesimal formation \citep{Goodman2000,Youdin2005}.      
 
 The continuum equations and the relevant parameters describing the evolution of dust-gas mixtures are given in Section~\ref{sec:mixtures}. Section~\ref{sec:SPHmixtures} describes the two-fluid SPH algorithm, addressing issues 1)-3). The code is benchmarked against a suite of test problems that we have specifically designed in order to provide standardised benchmarks for other two-fluid gas/dust codes, addressing issues 4) and 7) (Sec.~\ref{sec:tests}). The implicit timestepping scheme and treatment of non-linear drag (issues 5 and 6) are discussed in Paper~II. 

\section{Two-fluid mixtures in SPH}
\label{sec:SPHmixtures}

\subsection{Two-fluid gas and dust mixtures}
\label{sec:mixtures}

\subsubsection{Densities}
 The fact that dust grains of finite size occupy a finite volume is accounted for by defining the volume fraction available to the gas according to \citep[e.g.][]{Marble1970,harlow1975}
\begin{equation}
\theta = 1 - \frac{ \hrhod }{ \rhod }.
\label{eq:theta}
\end{equation}
This means that the volume densities of gas and dust $\hrhog$ and $\hrhod$, respectively, are distinguished from the intrinsic densities denoted $\rhog$ and $\rhod$, respectively, according to
\begin{eqnarray}
\hrhod & = & (1 - \theta) \rhod, \\
\hrhog & = & \theta \rhog.
\end{eqnarray}
The effects associated with finite dust particle size are mostly negligible in astrophysical problems since typically the intrinsic dust density $\rhod$ is much higher than the volume density $\hrhod$ and thus $\theta \approx 1$. We retain these terms, as in earlier SPH formulations \citep[c.f.][]{Monaghan1995} in order to retain a general algorithm that can be applied both within and outside of astrophysics.

The conservation of mass in a two-fluid mixture is thus expressed by the continuity equations
\begin{eqnarray}
\frac{\partial \hrhog}{\partial t} + \nabla . \left ( \hrhog \vg \right) & = & 0 \label{eq:mass_gas},\\
\frac{\partial \hrhod}{\partial t} + \nabla . \left ( \hrhod \vd \right) & = & 0 \label{eq:mass_dust},
\end{eqnarray}
where ${\bf v}_{\rm g}$ and ${\bf v}_{\rm d}$ are the gas and dust fluid velocities, respectively.

\subsubsection{Equations of motion}
 The equations of motion, expressing momentum conservation in a continuous, inviscid, two-fluid mixture of gas and dust are given by 
\begin{eqnarray}
\hrhog \left( \frac{\partial \vg}{\partial t} + \vg . \nabla \vg \right)  & = & - \theta\phantom{.}\nabla P_{\rm g} + \hrhog \textbf{f} -  \fp \label{eq:momentum_gas},\\[1em]
\hrhod \left( \frac{\partial \vd}{\partial t} + \vd . \nabla \vd \right) & = &  - \nabla P_{\rm d}  - \left(1 -\theta \right) \nabla P_{\rm g} + \hrhod \textbf{f} +  \fp \label{eq:momentum_dust},
\end{eqnarray}
where  $P_{\rm g}$ and $P_{\rm d}$ are the intrinsic pressures. Any intrinsic viscosities have been neglected. For astrophysical purposes it may be assumed that the dust is pressureless, i.e. $P_{\rm d} = 0$. Similarly, the term $ \left(1 -\theta \right)  \nabla P_{\rm g}$ in the momentum equation for the dust phase --- a buoyancy term related to the finite size of the dust particles --- is in general negligibly small. The reader should note that the definitions of physical quantities in a two fluid medium require the local fluid volume over which the averaging is performed to be defined \citep[see, e.g.][]{Marble1970,FanZhu}.

 The two fluids exchange momentum $\fp$, the drag force per unit volume, the expression for which is obtained by averaging the \textit{local} drag stress tensor (denoted $\mathbf{\epsilon}^{ij}_{\rm drag}$) over the surface area of the dust grains:
\begin{equation}
F_{\rm drag}^{{\rm V},i} = \frac{1}{V} \int_{A_{\mathrm{d}}} \mathbf{\epsilon}^{ij}_{\rm drag} \mathrm{d} A^{j} .
\label{eq:local_average}
\end{equation}
 In the case where the local distribution of dust particles is homogeneous (i.e., dust particles have the same mass, size and intrinsic density), Eq.~\ref{eq:local_average} simplifies to
\begin{equation}
\fp = K  (\vg - \vd).
\label{eq:def_fp}
\end{equation}
Note that since $\fp$ is a force per unit volume, the drag coefficient $K$, has dimensions of mass per unit volume per unit time. This coefficient is related to the drag coefficient on a single grain (denoted $K_{\rm s}$) by
\begin{equation}
K = \frac{\hrhod}{\md} K_{\rm s}
\end{equation}
where $\md$ is the mass per grain. The drag force (not per unit volume) on a single grain is given by
\begin{equation}
{\bf F}_{\rm drag} = K_{\rm s} (\vg - \vd).
\end{equation}
 In general $K$ (or equivalently $K_{s}$) can itself be a function of the relative velocity between the two fluids $\Delta v \equiv \vert \vg - \vd \vert$, resulting in a non-linear drag regime. In this, Paper~I we consider the simplest case of linear drag, where $K$ is constant with respect to $\Delta v$. Extension of our scheme to the main non-linear regimes applicable to astrophysics are considered in Paper~II.

 Finally, it should be noted that in general additional forces (e.g. the carried mass, Basset and Saffman forces, \citealt{FanZhu}) may be present in two-fluid systems. We have assumed in adopting Eqs.~\ref{eq:momentum_gas}--\ref{eq:momentum_dust} that these forces can be neglected for astrophysical applications.

\begin{figure*}
   \centering
   \includegraphics[angle=0, width=0.4\textwidth]{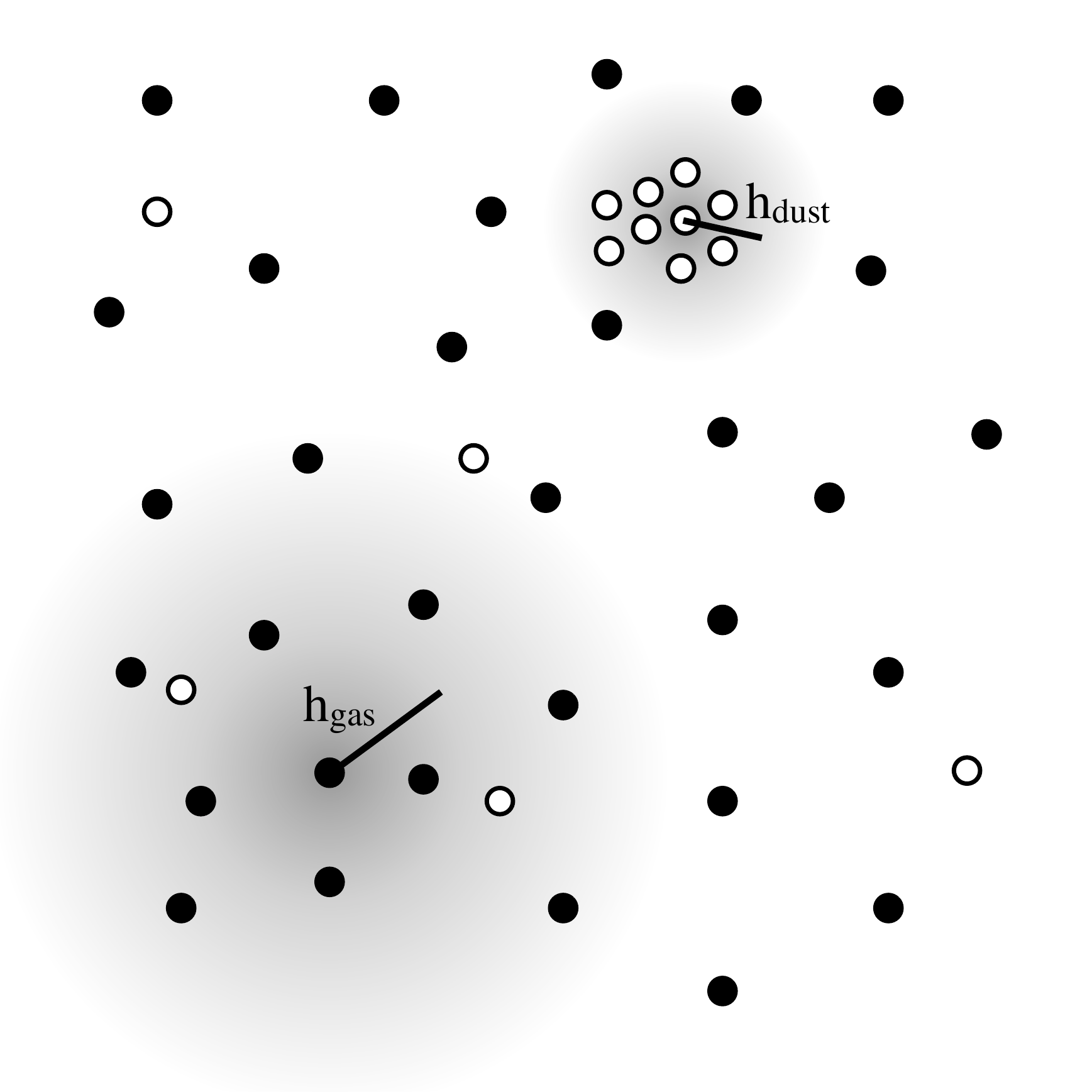} 
   \includegraphics[angle=0, width=0.4\textwidth]{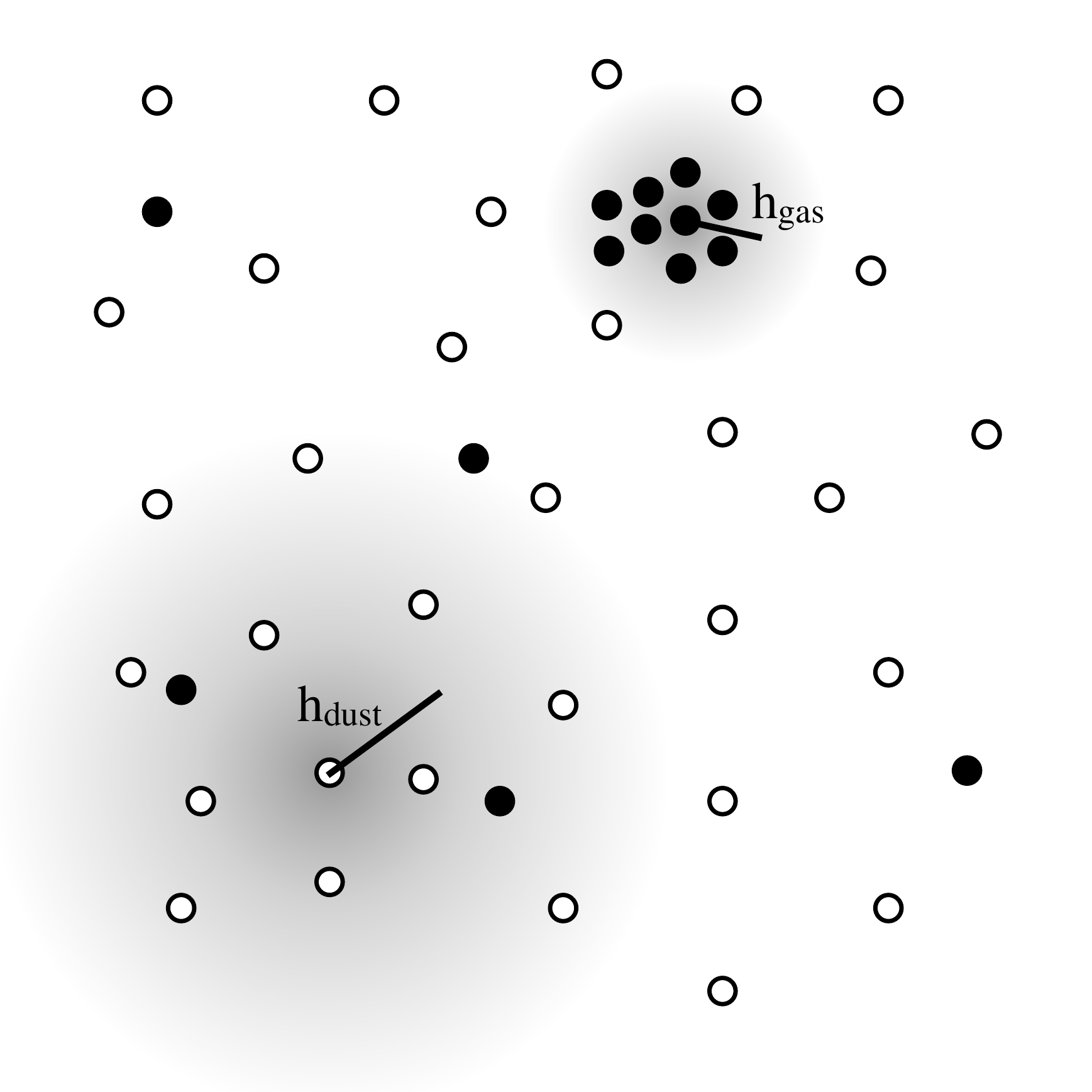} 
   \caption{Computing density in SPH gas (solid points) and dust (hollow circles) mixtures. Standard bell-shaped, Gaussian-like, kernels are adopted (weighting indicated by the shading), with a single smoothing length on each particle related to the local number density of particles of the \emph{same type}. This provides good density estimates in both extremes --- where dust is concentrated below the gas scale (left panel) and where gas is concentrated below the dust scale (right panel). The density of another fluid at the position of a reference fluid (e.g. dust density at the location of a gas particle) is computed using the same smoothing length but only neighbours of the desired type. This density is thus allowed to be identically zero, as would be the case for the density of gas-at-dust in the left panel (top), or dust-at-gas in the right panel.}
   \label{fig:dust-gas-smoothing}
\end{figure*}

\subsubsection{Energy equation}
The evolution equation for the specific internal energy of the gas, $u_{\rm g}$, is given by
\begin{equation}
\hrhog \frac{{\rm d}u_{\rm g}}{{\rm d}t} = -\Pg \left[ \theta \nabla\cdot{\bf v}_{\rm g} + (1 - \theta) \nabla\cdot{\bf v}_{\rm d} \right] + \Lambda_{\rm drag} + \Lambda_{\rm therm},
\label{eq:dudt}
\end{equation}
where the first term corresponds to the usual compressive ($P{\rm d}V$) term with the volume reduced by the dust filling factor $\theta$. The second term is the work done by the gas in triggering buoyancy effects. The third term is the frictional heating due to the drag force, given by
\begin{equation}
\Lambda_{\rm drag} = \hrhog K (\vg - \vd)^{2}.
\label{eq:dragheat}
\end{equation}
The fourth, thermal coupling, term arises when the internal temperature of the grains differs from the gas temperature \citep[c.f.][]{Marble1970,harlow1975}, and in general consists of terms related to heat transfer due to conduction ($\Lambda_{\rm cond}$) and radiation ($\Lambda_{\rm rad}$), given by
\begin{equation}
\Lambda_{\rm therm} \equiv \Lambda_{\rm cond} +  \Lambda_{\rm rad} = Q (T_{\rm g} - T_{\rm d}) + R (aT_{\rm g}^{4} - aT_{\rm d}^{4}),
\end{equation}
where $T_{\rm g}$ and $T_{\rm d}$ are the temperatures of the gas and dust, respectively, $a$ is the radiation constant and $Q$ and $R$ are coefficients, dependent on gas and dust properties, that characterise the heat transfer. The thermal energy of the dust evolves according to
\begin{equation}
\hrhod\frac{{\rm d}u_{\rm d}}{{\rm d}t} = -\Lambda_{\rm therm}.
\end{equation}

\subsection{Densities for two-fluid mixtures in SPH}
\label{sec:Densities}

\subsubsection{Computing densities in two-fluid SPH}
\label{sec:SPHdensities}
 For two-fluid mixtures, we require a density estimate \emph{for each phase}, corresponding to the exact solution of Eqs.~\ref{eq:mass_gas} and \ref{eq:mass_dust} in SPH. The main complication arises from the fact that the local particle spacing can be different for each fluid, implying that the two fluids should have different resolution lengths calculated based on the local particle number density of their own type. Figure~\ref{fig:dust-gas-smoothing} illustrates the two limiting cases, i.e. a high concentration of dust in a diluted gas (left panel) and conversely a high concentration of gas in a low density fluid of dust (right panel). In each case the smoothing length for each type is determined by the local number density of particles of \emph{the same type}. That is, the SPH translation of Eqs.~\ref{eq:mass_gas} and \ref{eq:mass_dust} correspond to
\begin{eqnarray}
\hrhoga = \sum_{b} m_{b}  W_{ab} (h_{a}); &\hspace{1cm} & h_{a} = \eta \left(\frac{\ma}{\hrhoga} \right )^{1/\nu}, \label{eq:h_density_gas} \\
\hrhoda = \sum_{j} m_{j}  W_{ij} (h_{i});  &\hspace{1cm}& h_{i} = \eta \left(\frac{m_{j}}{\hrhoda} \right )^{1/\nu},
\label{eq:h_density_dust}
\end{eqnarray}
where $\nu$ is the number of spatial dimensions and $\eta$ is a constant determining the resolution length as a function of the local particle spacing (typically $\eta = 1.2$ is a good choice for the standard cubic spline kernel, see \citealt{Price2011}). We adopt the convention that the indices $a,b,c$ refer to quantities computed on gas particles while $i,j,k$ refer to quantities computed on dust particles. Note that the densities and smoothing lengths are independently computed for each fluid and are thus --- so far --- only defined on particles of the same type. The numerical solution of Eqs.~\ref{eq:h_density_gas} and \ref{eq:h_density_dust} involves determining both $\hat{\rho}$ and $h$ for each type simultaneously, since they are mutually dependent, thus requiring an iterative procedure. The procedure is identical to that adopted in standard variable smoothing length SPH formulations (see e.g. \citealt{pm07} for details).

 An additional complication arises from the need to compute the volume filling fraction $\theta$ (Eq.~\ref{eq:theta}), defined on a \emph{gas} particle, $a$, according to
\begin{equation}
\thetaa = 1 - \frac{ \hat{\rho}_{{\rm d},a} }{ \rho_{\rm d} },
\label{eq:SPH_theta}
\end{equation}
which depends on the density of \emph{dust} at the gas particle location. Initially we considered computing this density using a second smoothing length for each particle based on neighbours of the \emph{other} type (this would in turn lead to multiple smoothing lengths on each particle if more types were present). However, the key point, illustrated by the right hand panel of Fig.~\ref{fig:dust-gas-smoothing}, is that it should be possible for the local density of dust at the gas location to be identically zero (giving $\theta = 1$) if no dust particles are found within the kernel radius computed with the gas smoothing length. Thus, the density of dust-at-gas should be calculated according to
\begin{equation}
\hat{\rho}_{{\rm d},a} = \sum^{N_{neigh, dust}}_{j=1} m_{j} W_{aj} (h_{a}),
\label{eq:hdensitydustatgas}
\end{equation}
where $h_{a}$ is the smoothing length of the \emph{gas} particle computed using gas neighbours as in Eq.~\ref{eq:h_density_gas}. In the case where no dust neighbours fall within the kernel radius, $\rho_{{\rm d},a} = 0$. This is a very simple and efficient method that can easily be generalised to multiple fluids, requires only one smoothing length per particle and does not require any significant additional computational expense.

 The discussion above resolves the first issue highlighted in Sec.~\ref{sec:intro}, namely how to deal with variable resolution in multi-fluid SPH, generalising the earlier fixed-smoothing-length formulation of \citet{Monaghan1995}. A similar discussion to the above applies to gravitational force softening on multiple fluids in $N$-body/SPH codes where the softening formulation is derived from a kernel density estimate \citep{pm07}, in particular for the case of a mixture of dark matter and baryonic gas \citep[e.g.][]{Merlin2010,id2011}.

\subsubsection{Kernel function}
The kernel function itself can be written as a function of the smoothing length $h$ and the dimensionless variable $q=\vert {\bf r} - {\bf r}' \vert / h$ in the form
\begin{equation}
W\left( r,h\right) = \frac{\sigma}{h^{\nu}} f\left( q \right)  ,
\label{eq:Kerform}
\end{equation}
where $\sigma$ is a normalisation constant. The standard Gaussian kernel is given by
\begin{equation}
f(q) = e^{-q^{2}},
\end{equation}
where $\sigma = \pi^{-\nu/2}$. The Gaussian is infinitely smooth (differentiable) but has the practical disadvantage of infinite range. A standard alternative (providing a Gaussian-like kernel but truncated at $2h$) is the $M_{4}$ cubic spline kernel \citep{monaghan92}
\begin{equation}
f(q) =
\begin{cases}
1 - \frac{3}{2}q^{2} + \frac{3}{4}q^{3} , & 0 \leq q < 1 ; \\
 \frac{1}{4}\left( 2 - q\right) ^{3},& 1 \leq q < 2 ;\\
0, & q \geq 2,
\end{cases}
\label{eq:spline}
\end{equation}
where $\sigma = \left[ 2/3 , 10/\left( 7\pi \right) , 1/ \pi\right]  $ in $ \left[1,2,3 \right] $ dimensions. An error analysis of the SPH density estimate \citep[e.g.][]{Price2011} shows that in general the measure of a good density kernel is that the normalisation condition
\begin{equation}
\sum_{b} \frac{m_{b}}{\rho_{b}} W_{ab} \approx 1,
\label{eq:Wnorm}
\end{equation}
is well satisfied for typical SPH particle distributions, corresponding to $\int W {\rm d}V = 1$ in the continuum limit. In general most bell-shaped (Gaussian-like) kernels, such as the cubic spline, fulfil this criterion \citep{fq96}.

 More accurate density estimates can be obtained --- at the price of additional computational expense --- by using kernels with extended range that form a better approximation to the Gaussian \citep[see][]{Price2011}. In particular the $M_{6}$ quintic kernel, truncated at $3h$, gives results that are in practice largely indistinguishable from the Gaussian, with the functional form
\begin{equation}
f(q) = 
\begin{cases}
(3-q)^{5} - 6(2-q)^{5} + 15(1-q)^{5}, & \text{$0 \leq q < 1$;} \\
(3-q)^{5} - 6(2-q)^{5}, & \text{$1 \leq q < 2$;} \\
(3-q)^{5}, & \text{$2 \leq q < 3$;} \\
0,                & \text{$q \geq 3$}.
\end{cases}
\label{eq:quintic}
\end{equation}
where $\sigma = [1/24, 96/(1199\pi), 1/(20\pi)]$. Use of the quintic is a factor of $(3/2)^{3} \approx 3.4$ times more expensive than the cubic spline (or other $2h$-truncated kernels) in three dimensions.

 The functional form of the $M_{4}$ cubic and $M_{6}$ quintic spline kernels are shown in the top row of Fig.~\ref{fig:kernels}, showing the kernel function $f(q)$ (solid/black lines) and its first (dashed/red lines) and second (short dashed/green line) derivatives.

\subsection{Equations of motion}
\label{sec:Without}
 As discussed by \citet{Price2011}, specifying the manner in which the density is calculated in SPH can be used to self-consistently determine the equations of motion and energy from a variational principle, using only the additional constraint of the first law of thermodynamics. For a two-fluid system, only the dissipationless part of the algorithm can be derived in this manner --- that is, not including the drag terms.

\subsubsection{Lagrangian}
  For a system consisting of gas and dust, the Lagrangian is given by
\begin{equation}
L = \sum_{b} \mb \left[ \frac{1}{2} \vb ^{2} - \ugb \left(\rhogb, \sgb \right) \right] + \sum_{k} \mk \left( \frac{1}{2} \vk ^{2} \right)
\label{eq:SPH_lagrangian}
\end{equation}
where $u_{b}$ is the thermal energy per unit mass of the gas (in general a function of the entropy $s$ and \emph{intrinsic} density $\rho$). The equations of motion can be derived from the Euler-Lagrange equations,
\begin{equation}
\frac{\mathrm{d}}{\mathrm{d}t} \left( \frac{\partial L  }{ \partial {\bf v}  } \right) = \frac{ \partial L }{ \partial {\bf r} } .
\label{eq:SPH_EL}
\end{equation}

\subsubsection{Equations of motion for the gas}
We first consider the evolution of the gas particles. The partial derivative of the Lagrangian with respect to the velocity $\va$ of a given gas particle $a$ provides:
\begin{equation}
\frac{\mathrm{d}}{\mathrm{d}t} \left( \frac{\partial L}{ \partial \va} \right) = \ma \frac{\mathrm{d} \va}{\mathrm{d}t} .
\label{eq:diff_vit_gas}
\end{equation}
The partial derivative of the Lagrangian with respect to the position $\ra$ of the gas particle $a$ is given by
\begin{equation}
 \frac{\partial L}{ \partial \ra} =  -\sum_{b} \mb \left. \frac{ \partial u_{b}  }{ \partial \rhogb }\right\vert_{s}  \frac{ \partial \rhogb  }{ \partial \ra },
\label{eq:diff_pos_gas1}
\end{equation}
where the entropy is constant for a non-dissipative system. Eq.~\ref{eq:diff_pos_gas1} differs from the usual expression for single fluids because of the distinction between the intrinsic and volume density of the gas caused by the finite volume occupied by dust. That is, the thermal energy depends on the \emph{intrinsic} density rather than the volume density, giving
\begin{equation}
\left.\frac{ \partial  \ugb}{\partial \rhogb  } \right\vert_{s}  = \frac{\Pgb}{\rhogb^{2}} =  \frac{\theta_{b}^{2} \Pgb}{\hrhogb^{2}}.
\end{equation}
The derivative of the intrinsic density with respect to the particle coordinates is given by
\begin{equation}
 \frac{ \partial \rhogb  }{ \partial \ra } = \left. \frac{ \partial \rhogb }{\partial \hrhogb } \right\vert_{\thetab} \frac{ \partial \hrhogb }{\partial \ra }  + \left. \frac{ \partial \rhogb}{\partial \thetab} \right\vert_{\hrhogb}  \frac{ \partial \thetab }{\partial \ra },
\end{equation}
where, from \ref{eq:SPH_theta}, we have
\begin{equation}
\left. \frac{ \partial \rhogb }{\partial \hrhogb } \right\vert_{\thetab} = \dst \frac{1}{\thetab}; \hspace{1cm} \left. \frac{ \partial \rhogb}{\partial \thetab} \right\vert_{\hrhogb}  = -\frac{\hrhogb}{\thetab^{2}}.
\label{eq:rel_SPH}
\end{equation}
The spatial derivative of the density sum for the gas (Eq.~\ref{eq:h_density_gas}) is given by
\begin{equation}
\frac{ \partial \hrhogb  }{ \partial \ra } = \frac{1}{\Omega_{b}} \sum_{c} \mc \left(\delta_{ba} - \delta_{ca} \right) \nabla_{a} W_{bc} (h_{b}),
\label{eq:diff_pos_hrho}
\end{equation}
where $\Omega$ is the usual variable smoothing length term
\begin{equation}
\Omega_{b} \equiv 1 -  \frac{\partial \hb }{\partial \hrhogb} \sum_{c} \mc \frac{\partial \kn{b}{c}  \dohb }{\partial \hb}.
\end{equation}
The spatial derivative of the volume filling fraction $\theta$ is given, from Eq.~\ref{eq:SPH_theta} by
\begin{equation}
\frac{\partial \theta_{b}}{\partial \ra} = - \frac{1}{\rho_{\rm d} }\frac{\partial \hat{\rho}_{{\rm d},b}}{\partial \ra},
\end{equation}
where
\begin{align}
\frac{\partial \hat{\rho}_{{\rm d},b}}{\partial \ra} = &  \sum_{j} \mj \left(\delta_{ba} - \delta_{ja} \right) \nabla_{a} W_{bj} (h_{b}) \nonumber \\
& + \frac{1 - \omdb}{\Omega_{b}}  \sum_{c} \mc \left(\delta_{ba} - \delta_{ca} \right) \nabla_{a} W_{bc} (h_{b}),
\end{align}
where $\Omega_{\rm d}$ is $\Omega$ computed only using dust particle neighbours, i.e.
\begin{equation}
\omdb = 1 - \frac{\partial \hb }{\partial \hrhodb} \sum_{j} \mj \frac{\partial \kn{b}{j}  \dohb}{\partial \hb}.
\label{eq:omegadb}
\end{equation}

Collecting Eqs.~\ref{eq:diff_pos_gas1}--\ref{eq:omegadb}, noting that $\delta_{ja} = 0$ (since a gas and dust index can never refer to the same particle) and using the fact that $\gkn{a}{b}{c} = -\gkn{a}{c}{b}$, gives
\begin{align}
 \frac{ \partial L  }{ \partial \ra }  = & - \ma \sum_{b} \mb \left[ \frac{\Pga \tilde{\theta}_{a}}{\omga \hrhoga^{2}} \gkn{a}{a}{b}\doha + \frac{\Pgb \tilde{\theta}_{b}}{\omgb \hrhogb^{2}} \gkn{a}{a}{b}\dohb \right] \nonumber\\  
 & - \ma \sum_{j} \mj \frac{\Pga \left(1 - \thetaa \right )}{\hrhoga \hat{\rho}_{{\rm d},a}} \gkn{a}{a}{j} \doha \label{eq:diff_pos_gas_2},
\end{align}
where we have defined $\tilde{\theta}$ to include the correction terms for a variable smoothing length, i.e.
\begin{equation}
\tilde{\theta} \equiv \theta + \frac{\hat{\rho}_{\rm g}}{\hrhod} ( 1 - \theta) (1 - \Omega_{\rm d}).
\end{equation}
Although this correction is necessary for strict energy conservation, it is expected to be negligibly small in practice, since $(1-\theta)$ is negligible for small grains and $(1-\Omega_{\rm d})$ is $\mathcal{O}(h^{2})$. Finally, the equations of motion for a gas particle, from the Euler-Lagrange equations, are given by
\begin{align}
 \frac{\mathrm{d} \va}{\mathrm{d}t}  = &  - \sum_{b} \mb \left[ \frac{\Pga \tilde{\theta}_{a}}{\omga\hrhoga^{2}} \gkn{a}{a}{b} \doha + \frac{\Pgb \tilde{\theta}_{b}}{\omgb\hrhogb^{2}} \gkn{a}{a}{b} \dohb \right]  \nonumber \\ 
 & - \sum_{j} \mj \frac{\Pga \left(1 - \thetaa \right )}{\hrhoga  \hat{\rho}_{{\rm d},a}} \gkn{a}{a}{j} \doha \label{eq:EL_gas}.
\end{align}
The reader should note that while the first term is a summation over gas particle neighbours, the second is summed over dust particle neighbours. Eq.~\ref{eq:EL_gas} may be straightforwardly shown to be a direct translation of Eq.~\ref{eq:momentum_gas} into SPH form. Note that the summation over dust particles (the buoyancy term) does not involve $\Omega$ since the smoothing length is independent of the dust particle positions.

\subsubsection{Equations of motion for the dust}
The partial derivative of the Lagrangian with respect to the velocity ${\bf v}_{i}$ of a given dust particle $i$ gives
\begin{equation}
\frac{\mathrm{d}}{\mathrm{d}t} \left( \frac{\partial L}{ \partial {\bf v}_{i}} \right) = m_{i} \frac{\mathrm{d} {\bf v}_{i}}{\mathrm{d}t}.
\label{eq:diff_vit_dust}
\end{equation}

 A buoyancy term arises in the dust because of the dependence of the gas internal energy on $\theta$, which in turn depends on the positions of dust particles. That is, 
\begin{equation}
 \frac{\partial L}{ \partial {\bf r}_{i}} =  -\sum_{b} \mb \left. \frac{ \partial u_{b}  }{ \partial \rhogb }\right\vert_{s}  \frac{ \partial \rhogb  }{ \partial {\bf r}_{i} },
\label{eq:diff_pos_dust1}
\end{equation}
where
\begin{equation}
 \frac{ \partial \rhogb  }{ \partial {\bf r}_{i}} =  \frac{ \partial \rhogb}{\partial \thetab}  \frac{ \partial \thetab }{\partial {\bf r}_{i} },
\end{equation}
and in turn,
\begin{align}
\frac{\partial \theta_{b}}{\partial {\bf r}_{i}} = & - \frac{1}{\rho_{\rm d} }\frac{\partial \hat{\rho}_{{\rm d},b}}{\partial  {\bf r}_{i}} \nonumber \\
= & - \frac{1}{\rho_{\rm d} } \sum_{j} \mj \left(\delta_{bi} - \delta_{ji} \right) \nabla_{i} W_{bj} (h_{b}) \nonumber \\
& - \frac{1 - \omdb}{\Omega_{b}} \sum_{c} \mc \left(\delta_{bi} - \delta_{ci} \right) \nabla_{i} W_{bc} (h_{b})
\label{eq:dthetadust}
\end{align}
 Collecting Eqs.~\ref{eq:diff_vit_gas}--\ref{eq:dthetadust} and noting that $\delta_{bi} = \delta_{ci} = 0$, we obtain the equations of motion for a dust particle in the form
\begin{equation}
\frac{\mathrm{d} {\bf v}_{i}}{\mathrm{d}t} = \sum_{b} \mb \frac{ \Pgb \left(1 - \thetab \right) }{\hrhogb \hrhodb } \gkn{i}{b}{i} \dohb .
\label{eq:EL_dust}
\end{equation}
where we have written the kernel using $\gkn{i}{i}{b} = -\gkn{i}{b}{i}$ to show that the force is equal and opposite to that in the gas (Eq.~\ref{eq:EL_gas}). It may be straightforwardly verified that Eq.~\ref{eq:EL_dust} is indeed a direct translation of Eq.~\ref{eq:momentum_dust} in SPH form.

 Equations~\ref{eq:EL_gas} and \ref{eq:EL_dust} may be combined to show that the total momentum is exactly conserved, i.e.
\begin{equation}
\frac{\rm d}{{\rm d}t} \left( \sum_{a} m_{a} {\bf v}_{a} +  \sum_{i} m_{i} {\bf v}_{i} \right) = 0.
\end{equation}

\subsubsection{Internal energy equation for the gas}
 The SPH form of the non-dissipative terms in the internal energy equation for the gas (Eq.~\ref{eq:dudt}) can similarly be derived from the SPH density estimates. In the absence of dissipation the evolution equation for a given gas particle $a$ is given by
\begin{equation}
\frac{{\rm d}u_{a}}{{\rm d}t} = \frac{P_{a}}{\rho_{a}^{2}} \frac{{\rm d}\rho_{a}}{{\rm d}t} = \frac{P_{a}}{\rho_{a}^{2}} \left[ \left. \frac{ \partial \rhoga }{\partial \hrhoga } \right\vert_{\thetaa} \frac{{\rm d} \hrhoga }{{\rm d}t}  + \left. \frac{ \partial \rhoga}{\partial \thetaa} \right\vert_{\hrhoga}  \frac{{\rm d} \thetaa }{{\rm d}t } \right ].
\end{equation}
Using the expressions (\ref{eq:rel_SPH}) and simplifying using (\ref{eq:SPH_theta}), we have
\begin{equation}
\frac{{\rm d}u_{a}}{{\rm d}t} = \frac{\theta_{a} P_{a}}{\hat{\rho}_{a}^{2}}\frac{{\rm d} \hrhoga }{{\rm d}t} +  \frac{(1 - \theta_{a}) P_{a}}{\hrhoga \hat{\rho}_{{\rm d},a}} \frac{{\rm d}  \hat{\rho}_{{\rm d},a} }{{\rm d}t}.
\end{equation}
Taking the time derivative of the density sums (\ref{eq:h_density_gas}) and (\ref{eq:hdensitydustatgas}) we have
\begin{eqnarray}
\frac{{\rm d} \hrhoga }{{\rm d}t} & = & \frac{1}{\Omega_{a}} \sum_{b} m_{b} \left( \va - \vb \right) \cdot \nabla_{a} W_{ab} (h_{a}), \\
 \frac{{\rm d}  \hat{\rho}_{{\rm d},a} }{{\rm d}t} & = &  \sum^{N_{neigh, dust}}_{j=1} m_{j} \left( {\bf v}_{a} - {\bf v}_{j} \right)  \cdot \nabla_{a} W_{aj} (h_{a}), \nonumber \\
 & & 
+ \frac{(1 -  \Omega_{{\rm d}, a})}{\Omega_{a}}\sum_{b} m_{b} \left( \va - \vb \right) \cdot \nabla_{a} W_{ab} (h_{a})
\end{eqnarray}
giving the SPH internal energy equation in the form
\begin{align}
\frac{{\rm d}u_{a}}{{\rm d}t}  & = \frac{\tilde{\theta}_{a} P_{a}}{\Omega_{a} \hat{\rho}_{a}^{2} } \sum_{b} m_{b} \left( \va - \vb \right) \cdot \nabla_{a} W_{ab} (h_{a}) \nonumber \\ & +  \frac{(1 - \theta_{a}) P_{a}}{ \hrhoga \hat{\rho}_{{\rm d},a}} \sum^{N_{neigh, dust}}_{j=1} m_{j} \left( {\bf v}_{a} - {\bf v}_{j} \right) \cdot \nabla_{a} W_{aj} (h_{a}),
\end{align}
which indeed can be shown to be an SPH translation of the first two terms in Eq.~\ref{eq:dudt}.

\subsection{SPH representation of drag terms}
\label{sec:Drag}

\subsubsection{Drag interpolation}

 The remaining aspect is to provide an SPH representation of Eq.~\ref{eq:def_fp}, specifying the drag term $\fp$ involved in Eqs.~\ref{eq:momentum_gas}--\ref{eq:momentum_dust}. 
\citet{Monaghan1995} proposed an SPH interpolation of the drag term given by
\begin{equation}
\left< K \Delta \mathbf{v} \right> = \nu \int K\left(\mathbf{x} , \mathbf{x}' \right) \left\{ \left[ \vg ( \mathbf{x}) - \vd( \mathbf{x}' ) \right] \cdot \hat{\mathbf{r}} \right\} \hat{\mathbf{r}} D\left(\mathbf{x} - \mathbf{x}' ,h \right)  \mathrm{d} \mathbf{x}' , 
\label{eq:SPH_deltavD}
\end{equation}
where $\hat{\mathbf{r}}$ is the unit vector defined by:
\begin{equation}
\hat{\mathbf{r}} = \frac{\mathbf{x} - \mathbf{x}'}{\vert \mathbf{x} - \mathbf{x}' \vert} ,
\label{eq:def_rhat}
\end{equation}
and $\nu$ is the number of spatial dimensions of the system (and \emph{not} the inverse of the number of spatial dimensions as one might intuitively guess --- see below). \citet{Monaghan1995} proposed this formulation --- with velocity difference projected along the line of sight joining the particles --- mainly because it gives exact conservation of both linear and angular momentum in the resulting drag terms.

 As the SPH interpolation of the drag term does not come from the Euler-Lagrange equations derived for non-dissipative term form the SPH Lagrangian, the kernel function used in the drag term is not constrained to be the same function $W$ used for the density (as assumed by \citealt{Monaghan1995}). Indeed one of our findings from this paper (discussed below) is that use of a standard (bell-shaped) density kernel for drag computations can be significantly inaccurate. We thus use $D$ to denote the kernel employed for the drag interpolation.

\subsubsection{Choice of smoothing length in the drag terms}
\label{sec:choiceh}
 A key issue is the choice of smoothing length involved in the interpolation term (\ref{eq:SPH_deltavD}) when the gas and dust have different spatial resolutions, as illustrated in Fig.~\ref{fig:dust-gas-smoothing}. We have found from experiment that it is very important to smooth the drag term using the maximum smoothing length of the two fluids, rather than using an average (c.f. Sec.~\ref{sec:dustydisc} and also \citealt{Ayliffe2011}). Otherwise, unphysical resolution-dependent clumping of one fluid below the scale of the other can occur. For gas this presents less of a problem because there remain pressure gradients that prevent such clumping. However, for dust it is crucial since there are no forces that can otherwise counterbalance any artificial over-concentration. Since most astrophysical problems involve the concentration of dust in a flow of gas, a straightforward approach is to simply use the gas smoothing length when computing the drag interaction. Unless otherwise specified (Sec.~\ref{sec:dustydisc}) this is the approach we adopt in this paper.

 \subsubsection{Errors in the integral drag interpolant}
\label{sec:draginterpretation}

 The origin of Eq.~\ref{eq:SPH_deltavD} can be understood by considering the projection of $\left< \Delta \mathbf{v} \right>$ onto $\hat{\mathbf{r}}_{\alpha \alpha'}$, the projection of $\hat{\mathbf{r}}$ onto the coordinate $\alpha$ (which equivalently denotes the coordinates $x$, $y$ or $z$ as the system is invariant by rotation)  and use a Taylor expansion of $K$ and $\vd( \mathbf{x}')$ around their values on $\mathbf{x}$: 
\begin{align}
\left< K \Delta \mathbf{v} \right>^{\alpha} = & \nu \int \mathrm{d} \mathbf{x}'  \hat{\bf r}^{\alpha} \nonumber \\
& \left\lbrace  K \Delta \mathbf{v}(\mathbf{x}) + \frac{\partial (K \Delta \mathbf{v})}{\partial \mathbf{x}} \cdot (\mathbf{x} - \mathbf{x}' ) + \mathcal{O}\left((\mathbf{x} - \mathbf{x}' )^{2} \right) \right\rbrace \nonumber \\
& \cdot \hat{\mathbf{r}} D\left(\mathbf{x} - \mathbf{x}' , h\right) .
\end{align}
giving
\begin{equation}
\left< K \Delta \mathbf{v} \right>^{\alpha} = \nu K \Delta \mathbf{v}(\mathbf{x})^{\beta}  I^{\alpha\beta} + \nu \frac{\partial (K \Delta \mathbf{v}^{\alpha})}{\partial \mathbf{x}^{\gamma}} J^{\alpha\beta\gamma} + \mathcal{O}\left(h^{2} \right),
\label{eq:expand_SPH_drag}
\end{equation}
where
\begin{eqnarray}
I^{\alpha \beta} & \equiv & \int \mathrm{d} \mathbf{x}'  \hat{r}^{\alpha} ~ \hat{r}^{\beta} D\left(\mathbf{x} - \mathbf{x}' , h\right), \label{eq:SPH_integrals1} \\
J^{\alpha \beta \gamma} & \equiv & \int \mathrm{d} \mathbf{x}'  \hat{r}^{\alpha} \hat{r}^{\beta}  \hat{r}^{\gamma} D\left(\mathbf{x} - \mathbf{x}' , h\right).
\label{eq:SPH_integrals2}
\end{eqnarray}
This shows that Eq.~(\ref{eq:SPH_deltavD}) is a second-order approximation to the drag term, that is,
\begin{equation}
\left< K \Delta \mathbf{v} \right>^{\alpha}=  K \Delta \mathbf{v}^{\alpha} + \mathcal{O}\left(h^{2}\right),
\label{eq:approx_deltav}
\end{equation}
provided the normalisation conditions
\begin{eqnarray}
I^{\alpha \beta} & = & \frac{\delta^{\alpha\beta}}{\nu}, \label{eq:Inorm} \\
J^{\alpha \beta \gamma} & = & 0, \label{eq:Jnorm}
\end{eqnarray}
hold. Condition (\ref{eq:Jnorm}) and the zeroing of the off-diagonal terms in Eq.~\ref{eq:Inorm} may be proved straightforwardly by the fact that the integrals in (\ref{eq:SPH_integrals1})--(\ref{eq:SPH_integrals2}) are odd. The normalisation condition of the diagonal terms in Eq.~\ref{eq:Inorm} arises because in 3D we have
\begin{eqnarray}
I^{xx} + I^{yy} + I^{zz} & = & 1, \\
I^{xx} = I^{yy} = I^{zz}, & &
\end{eqnarray}
giving $I^{xx} = I^{yy} = I^{zz} = 1/\nu$. This explains the factor of $\nu$ in front of the drag summation term.

\subsubsection{Discretisation of drag term}

Discretising Eq.~\ref{eq:SPH_deltavD} provides the SPH translation of the acceleration due to the drag term for both the gas and the dust. Replacing the integral by a summation (over particles of the opposing type) and $\rho {\rm d}V$ with the particle mass, we have
\begin{equation}
 \left( \frac{\mathrm{d}\va}{\mathrm{d}t} \right) _{\mathrm{drag}} = \frac{1}{\hrhog}\left< K \Delta \mathbf{v} \right> =  \nu \sum_{j} m_{j} \frac{K_{aj}}{\hat{\rho}_{a} \hat{\rho}_{j}} \left({\bf v}_{aj} \cdot \hraj \right) \hraj D_{aj} (h_{a}) ,
\label{eq:SPH_drag_gas}
\end{equation}
for a gas particle and
\begin{equation}
 \left( \frac{\mathrm{d}{\bf v}_{i}}{\mathrm{d}t} \right) _{\mathrm{drag}} = \frac{1}{\hrhog} \left< K \Delta \mathbf{v} \right> = - \nu \sum_{b} \mb \frac{K_{bi}}{\hat{\rho}_{b} \hat{\rho}_{i}} \left({\bf v}_{bi} \cdot \hrbi \right) \hrbi D_{ib}  (h_{b}) ,
\label{eq:SPH_drag_dust}
\end{equation}
for a dust particle, where we have defined ${\bf v}_{aj} \equiv {\bf v}^{\rm g}_{a} - {\bf v}^{\rm d}_{j}$ and $\hat{\bf r}_{aj} \equiv ({\bf r}_{a} - {\bf r}_{j})/ \vert {\bf r}_{a} - {\bf r}_{j} \vert$. Importantly, from Eqs.~\ref{eq:SPH_drag_gas}--\ref{eq:SPH_drag_dust}, we have:
\begin{equation}
\sum_{a} \ma \left( \frac{\mathrm{d}\va}{\mathrm{d}t} \right) _{\mathrm{drag}} + \sum_{i} m_{i} \left( \frac{\mathrm{d}{\bf v}_{i}}{\mathrm{d}t} \right) _{\mathrm{drag}} = 0,
\label{eq:cons_mom_drag}
\end{equation}
which ensures that the momentum is exactly exchanged between the gas and the dust phase by the SPH formalism. Similarly
\begin{equation}
\sum_{a} \ma  {\bf r}_{a} \times \left(\frac{\mathrm{d}\va}{\mathrm{d}t} \right) _{\mathrm{drag}} + \sum_{i} m_{i}  {\bf r}_{i} \times \left( \frac{\mathrm{d}{\bf v}_{i}}{\mathrm{d}t} \right)_{\mathrm{drag}} = 0,
\label{eq:cons_angmom_drag}
\end{equation}
showing that the total angular momentum is conserved.

\subsubsection{Errors in the SPH drag interpolation}
A key point to note in the formulation of drag terms is that the criterion for an accurate kernel drag estimate is different from that required for an accurate density estimate (Eq.~\ref{eq:Wnorm}). Taking Eq.~\ref{eq:SPH_drag_gas} and expanding the velocities and the drag coefficient $K$ around the position of the gas particle ${\bf r}_{a}$, to lowest order, i.e.
\begin{equation}
K_{aj} \left( {\bf v}^{g}_{a}  - {\bf v}^{d}_{j} \right) = K_{a} \left( {\bf v}^{g}_{a} - {\bf v}^{d}_{a}  \right) + \mathcal{O}(h),
\end{equation}
we find
\begin{equation}
 - \nu \frac{K_{a}}{\hat{\rho}_{a}} \left( {\bf v}^{g}_{a} - {\bf v}^{d}_{a} \right) \cdot \sum_{j} \frac{m_{j}}{\hat{\rho}_{j}} \hraj \hraj D_{aj} + \mathcal{O}(h),
\end{equation}
implying a discrete normalisation condition on the drag kernel of the form
\begin{equation}
\nu \sum_{j} \frac{m_{j}}{\hat{\rho}_{j}} \hraj^{\alpha} \hraj^{\beta} D_{aj} \approx \delta^{\alpha \beta}.
 \label{eq:dragnorm}
\end{equation}
 The condition (\ref{eq:dragnorm}) implies that the summation on the diagonal terms ($xx$, $yy$, $zz$) are equal to unity, while the summations on off-diagonal terms ($xy$, $xz$, $yz$) should be zero. The accuracy with which this normalisation condition is satisfied depends on the particle arrangement. While we find that the diagonal terms are well computed using standard (bell-shaped) kernels, we find that --- apart from the special case where the dust particles lie on top of the gas particles --- the off-diagonal terms can be very poorly normalised. Fig.~\ref{fig:kernelerrors} shows the $xy$ component of Eq.~\ref{eq:dragnorm} as a function of the smoothing length (in units of the particle spacing, $\Delta x$) computed for a dust particle offset by $\Delta x/4$ in the x-direction from a cubic lattice of gas particles in 3D. Using the cubic spline kernel (top left) results in errors of order 5-10\% of the diagonal terms for reasonable neighbour numbers ($h/\Delta x \approx 1.1-1.5$). Furthermore, improving the smoothness of the kernel by using the $M_{6}$ quintic or even the Gaussian (top right) does \emph{not} significantly reduce the error. In numerical tests (Sec.~\ref{sec:dustybox}) this manifests as a large error in the drag between the two fluids, implying that a more suitable kernel is highly desirable.
 
 \begin{figure}
   \centering
   \includegraphics[angle=0, width=\columnwidth]{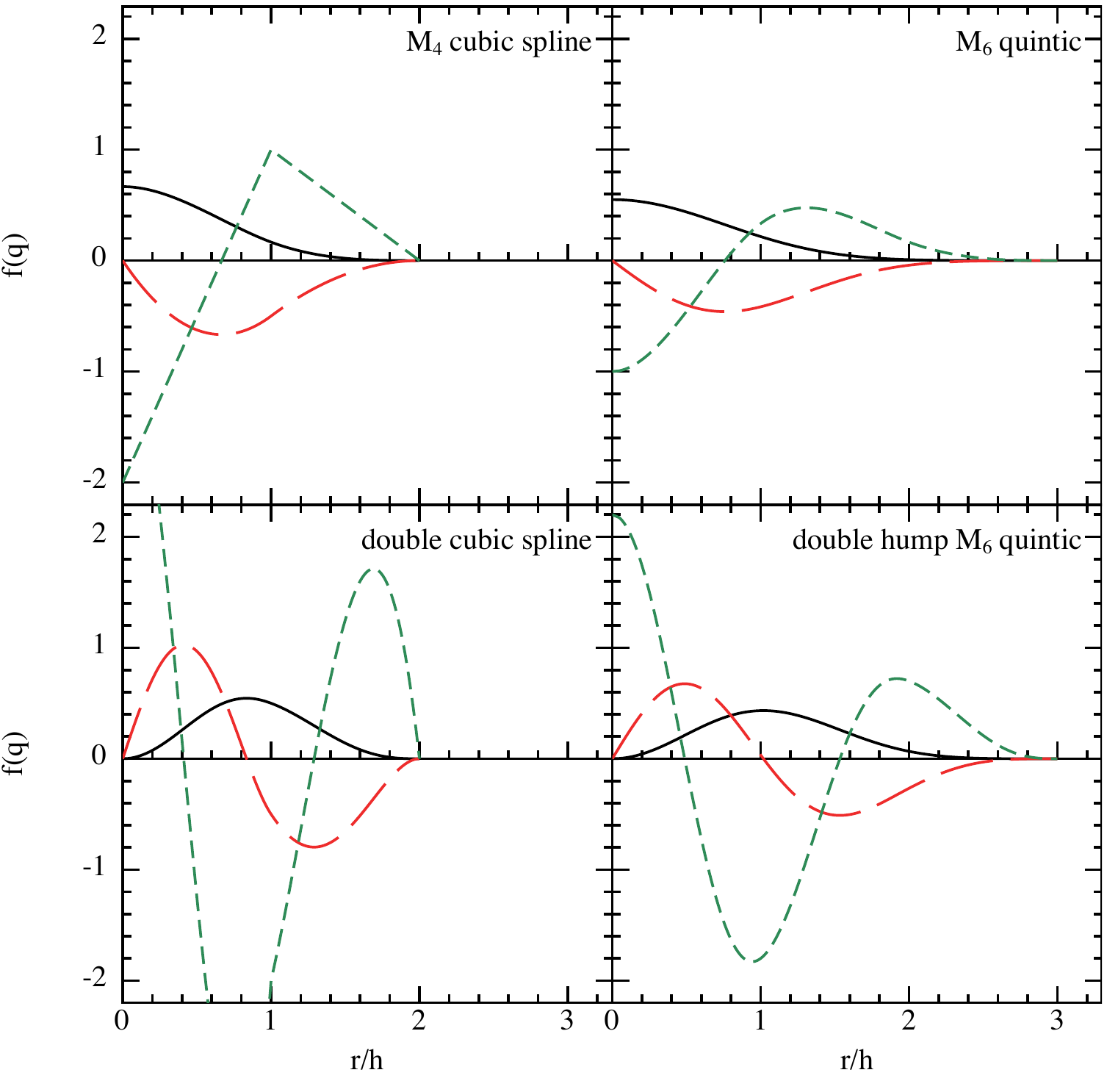} 
   \caption{Functional form of the standard bell-shaped cubic spline (top left) and quintic (top right) kernels, compared to the double-hump versions of these kernels (bottom row). Kernel functions are shown by the solid/black lines, while the long-dashed/red and short-dashed/green lines correspond to the first and second derivatives, respectively. We find that double-hump kernels are significantly more accurate than bell-shaped kernels when computing SPH drag terms (see Fig.~\ref{fig:kernelerrors}). }
   \label{fig:kernels}
\end{figure}

\begin{figure}
   \centering
   \includegraphics[angle=0, width=\columnwidth]{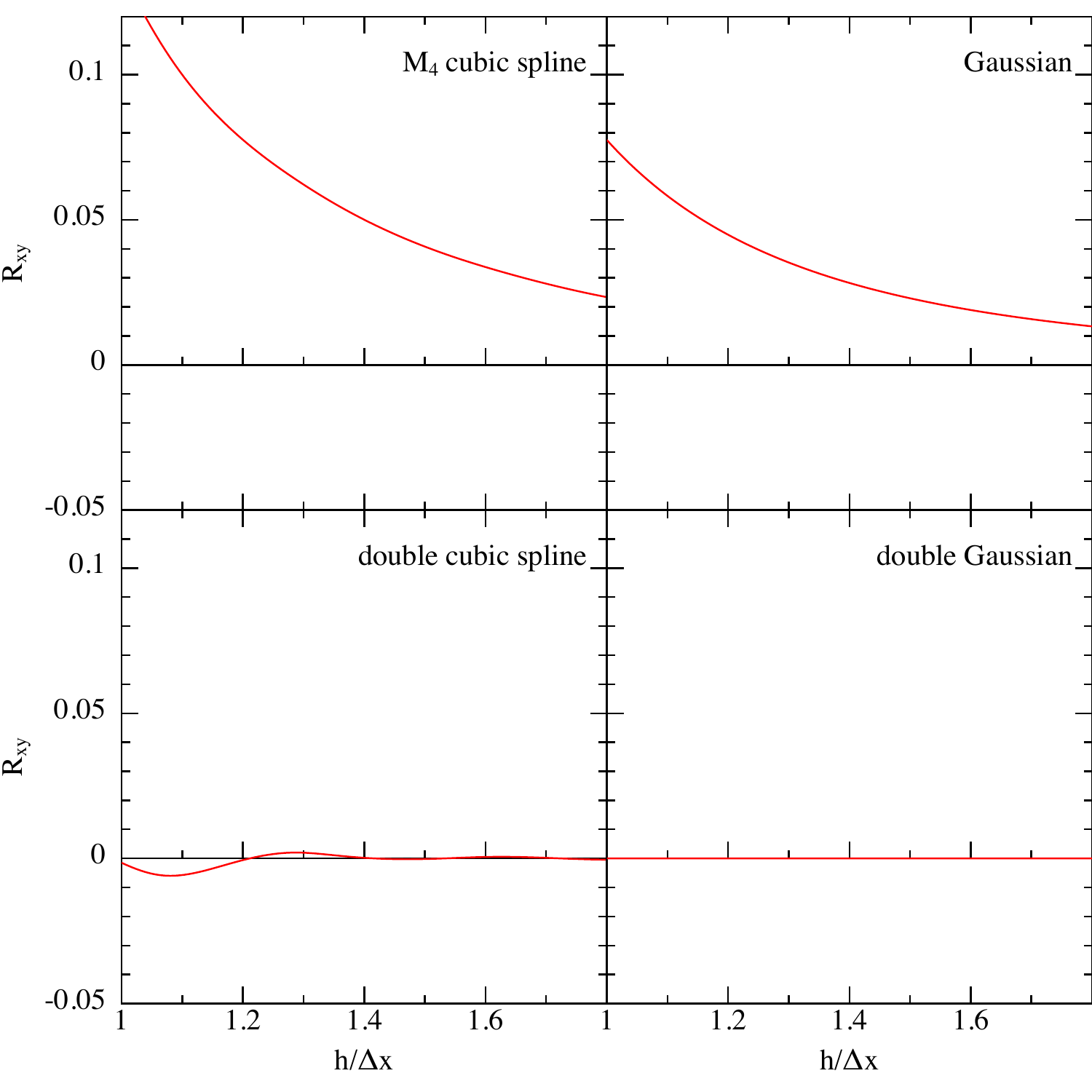} 
   \caption{Accuracy with which the normalisation condition for the drag force is computed using standard bell-shaped kernels (top row) and double-hump kernels (bottom row). The plots show the $xy$ component of Eq.~\ref{eq:dragnorm} computed on a dust particle offset from a regular cubic lattice of gas particles, as a function of the smoothing length in units of the particle spacing ($h/\Delta x$). With the bell-shaped kernels (top row) the errors are of order $5-10\%$ (the off-diagonal terms should sum to zero). Changing to double-hump shaped kernels (bottom row) gives errors $\lesssim 0.5\%$.}
   \label{fig:kernelerrors}
\end{figure}
 
\subsubsection{Drag kernel function}
\label{sec:dragkernel}
 After conducting a search for suitable alternative kernels, we found that the so-called ``double-hump'' shaped kernels \citep{fq96} gave a substantial improvement in accuracy --- that is, giving errors in the computation of Eq.~\ref{eq:dragnorm} of similar order to the bell-shaped kernels in computing Eq.~\ref{eq:Wnorm}. Defining the kernel function as previously
\begin{equation}
D\left( r,h\right) = \frac{\sigma}{h^{\nu}} g\left( q \right),
\label{eq:Kerformdrag}
\end{equation}
we construct double-hump kernels from the $M_{4}$ cubic and $M_{6}$ quintic kernels using
\begin{equation}
g(q) = q^{2} f(q),
\end{equation}
giving, for example, the ``double $M_{4}$ cubic'' (bottom left panel of Fig.~\ref{fig:kernels}), the ``double $M_{6}$ quintic'' (bottom right panel of Fig.~\ref{fig:kernels}) and similarly the ``double Gaussian''. The normalisation constants are found in the usual manner by enforcing $\int D {\rm d}V = 1$, i.e.,
\begin{equation}
\sigma \int g(q) {\rm d}V = 1,
\end{equation}
where ${\rm d}V$ corresponds to ${\rm d}q$, $2\pi q {\rm d}q$ and $4\pi q^{2} {\rm d}q$ in one, two and three dimensions, respectively. The normalisation constants for the double cubic, double quintic and double Gaussian are given by
\begin{eqnarray}
\sigma_{\textrm{double M}_{4}} & = & \left[2, \frac{70}{31\pi}, \frac{10}{9\pi} \right]; \\
\sigma_{\textrm{double M}_{6}} & = & \left[\frac{1}{60}, \frac{42}{2771\pi}, \frac{1}{168\pi} \right]; \\
\sigma_{\textrm{double Gaussian}} & = & \frac{2}{\nu} \pi^{-\nu/2},
\end{eqnarray}
in [1,2,3] dimensions. The computation of the off-diagonal term in (\ref{eq:dragnorm}) for the double-cubic and double-Gaussian kernels are shown in the bottom row of Fig.~\ref{fig:kernelerrors} and indeed show a substantial improvement, giving errors $\lesssim 0.5 \%$ compared to the $5-10\%$ errors obtained using the standard kernels (top row). This improvement in accuracy is also reflected in our numerical tests (c.f. Sec.~\ref{sec:dustybox}).

 It is also possible to physically understand the reason why the double hump kernel is suited to deal with drag computation. In treating multi-fluid interactions, one requires the information of one type of particle at the location of a particle of the opposing type. Assuming that the number of dimensions of the space is three, the SPH smoothing of the physical quantity $A$ corresponds approximately to
\begin{equation}
\begin{array}{rcl}
\dst 4 \pi^{2}  \!\! \int_{0}^{1} \!\!\! A\left(q \right) D(q) \mathrm{d}q  & \simeq & \dst 4 \pi^{2}  \!\! \int_{0}^{1} \!\!\! A\left(q \right) \frac{\delta\left(q + q_{\mathrm{M}} \right) + \delta \left(q - q_{\mathrm{M}} \right)}{2} \mathrm{d}q \\[1.3em]
&& = \dst \frac{A\left(- q_{\mathrm{M}} \right) + A\left(q_{\mathrm{M}} \right)}{2},
\end{array}
\label{eq:approx_hump}
\end{equation}
showing that --- for a given particle --- the double hump kernel provides an average value of a physical quantity stored in the neighbours of the other species and located at a distance $q = q_{\mathrm{M}}$ of the particle. For the same reason, the poor accuracy of the bell-shaped kernels can be understood because the maximum weight corresponds to $q = 0$, where in general, no particle of the other type is present.

\subsubsection{Frictional heating terms due to drag}

When the system is made of a single gas fluid, the \emph{specific} thermal energy $u$ is a function of state whose total derivative is expressed by:
\begin{equation}
\rm{d} u = T \rm{d} s + \frac{P}{\rho^{2}} \rm{d} \rho.
\label{eq:du_SPH}
\end{equation}
To generalise this relation with two-fluids interacting with a drag term, we derive an additional term for Eq.~\ref{eq:du_SPH} arising from exchange of momentum (all the other quantities fixed) considering a closed thermally isolated system made of gas and dust SPH particles, whose energy exchange arises only because of momentum exchange (i.e. drag) between two states (denoted \emph{i} and \emph{f}, respectively). Applying the first law of thermodynamics to an infinitesimal transformation of the system, we have:
\begin{equation}
\rm{d} u + \rm{d} e_{\rm{k}} = \delta w_{\rm{i} \to \rm{f}} +  \delta q_{\rm{i} \to \rm{f}},
\label{eq:first_principle}
\end{equation}
where $u$ is the total specific internal energy, $e_{\rm{k}}$ is the macroscopic kinetic energy of the system and $w$ is the total work and $q$ is the total heat exchanged during the transformation. Assuming that the transformation occurs slowly enough for the gas to remain in thermodynamic equilibrium, Eq.~\ref{eq:first_principle} reduces to:
\begin{equation}
\rm{d} u\vert_{s,\rho} + \rm{d} e_{\rm{k}} = T \rm{d} s + \frac{P}{\rho^{2}} \rm{d} \rho = 0.
\label{eq:first_principle_rev}
\end{equation}
Consequently,
\begin{align}
\rm{d} u\vert_{s,\rho} = - \rm{d} e_{\rm{k}} = & - \sum_{a} \frac{\left(\bf{v}_{a} + \left. \rm{d}\bf{v}_{a} \right\vert_{s_{a},\rho_{a}} \right)^{2}}{2} + \sum_{a} \frac{\bf{v}_{a}^{2}}{2} \\ \nonumber
& - \sum_{k} \frac{\left(\bf{v}_{k} + \left. \rm{d}\bf{v}_{k} \right\vert_{s_{k},\rho_{k}} \right)^{2}}{2} + \sum_{k} \frac{\bf{v}_{k}^{2}}{2} .
\label{eq:dek}
\end{align}
As the conservation of the momentum during the transformation ensures that:
\begin{equation}
\sum_{k} \left. \rm{d} \bf{v}_{k} \right\vert_{s_{k},\rho_{k}} = - \sum_{a} \left. \rm{d} \bf{v}_{a} \right\vert_{s_{a},\rho_{a}} ,
\end{equation}
we obtain:
\begin{equation}
\rm{d} u\vert_{s,\rho} = \sum_{a}  \bf{v}_{ka} \cdot \left. \rm{d} \bf{v}_{a} \right\vert_{s_{a},\rho_{a}}  .
\end{equation}
Using the expression Eq.~\ref{eq:SPH_drag_gas} which gives the evolution of the velocity of a gas particle due to drag term (i.e. at constant specific entropy and density):
\begin{equation}
\frac{\rm{d} u\vert_{s,\rho}}{\rm{d} t} = \sum_{a} \frac{\rm{d} u_{a}}{\rm{d} t}   = \sum_{a}  \mathbf{v}_{ka} \cdot \left[  \nu \sum_{k} \mk \frac{\kak}{\hat{\rho}_{a} \hat{\rho}_{k}} \left(\vak \cdot \hrak \right) \hrak D_{ak} (h_{a}) \right]  .
\label{eq:coms_mom_transfo}
\end{equation}
which implies that the evolution of the specific internal energy for each gas particle is given by:
\begin{equation}
\left( \frac{\rm{d} u_{a}}{\rm{d} t}\right)_{\rm drag} = \frac{\Lambda_{\rm drag}}{\hat{\rho}_{a}} =  \nu \sum_{k}   \mk \frac{\kak}{\hat{\rho}_{a} \hat{\rho}_{k}} \left(\vak \cdot \hrak \right)^{2} D_{ak} (h_{a}) .
\label{eq:evol_SPH_energy_drag}
\end{equation}
The positive value of ${\rm d} u_{a}/{\rm{d} t}$ (and the fact that the kernel function is always positive) ensures a positive definite contribution to the specific internal energy of each SPH gas particle from frictional drag heating. Eq.~\ref{eq:evol_SPH_energy_drag} provides the SPH translation of the drag heating term (Eq.~\ref{eq:dragheat}). It is straightforward to show that with the above expression the total energy is exactly conserved, i.e.
\begin{equation}
\sum_{a} m_{a} \frac{{\rm d}u_{a}}{{\rm d}t} + \sum_{a} m_{a} {\bf v}_{a}\cdot \frac{{\rm d}{\bf v}_{a}}{{\rm d}t} + \sum_{i} m_{i} {\bf v}_{i}\cdot \frac{{\rm d}{\bf v}_{i}}{{\rm d}t}= 0.
\end{equation}

\subsubsection{Thermal coupling terms}
The thermal coupling terms can be expressed in SPH form using \citep{Monaghan1995}
\begin{equation}
\Lambda_{{\rm therm}, a} = \sum_{j} m_{a} \frac{Q_{aj}}{\hat{\rho}_{a}\hat{\rho}_{j}} (T_{a} - T_{j}) W_{aj}  + \sum_{j} m_{a} \frac{R_{aj}}{\hat{\rho}_{a}\hat{\rho}_{j}} a (T_{a}^{4} - T_{j}^{4}) W_{aj},
\end{equation}
for a gas particle, and
\begin{equation}
\Lambda_{{\rm therm},i} = -\sum_{b} m_{i} \frac{Q_{bi}}{\hat{\rho}_{i}\hat{\rho}_{b}} (T_{b} - T_{j}) W_{bi}  - \sum_{b} m_{b} \frac{R_{bi}}{\hat{\rho}_{b}\hat{\rho}_{i}} a (T_{b}^{4} - T_{i}^{4}) W_{bi},
\end{equation}
for a dust particle. Note that we use the standard SPH kernel for the thermal coupling terms. Detailed study of the effect of the thermal coupling terms, or specific expressions for $Q$ and $R$, are beyond the scope of this paper, and thus for the tests in Sec.~\ref{sec:tests} we simply set $T_{\rm d}=T_{\rm g}$.

\subsubsection{Drag coefficient}
 In general the drag coefficient $K$ is a function of the properties of both the gas and dust. For example in the linear Epstein regime relevant to dilute gases in the limit of low Mach numbers, the coefficient $\kak$ is given by
\begin{equation}
\kak = \frac{4}{3} \pi   \sqrt{\frac{8}{\pi \gamma}} \frac{\hat{\rho}_{k}}{m_{\rm{d}}} \frac{\hrhoga}{\thetaa} s^{2}c_{\mathrm{s},a} ,
\label{eq:exprkak_Eps}
\end{equation}
where $s$ is the grain radius, $m_{\rm{d}}$ is the grain mass, $\gamma$ is the adiabatic index, $c_{\mathrm{s},a}$ is the gas sound speed. Since the basic dust-gas algorithm described above is insensitive to the specific form of the drag, we consider only constant drag coefficients in this paper in order to benchmark the method. The detailed implementation of a full range of both linear and non-linear physical drag formulations is considered in Paper~II.

\section{Timestepping}
\label{sec:timestepping}

\subsection{Empirical timestep criterion}
The drag terms impose an additional constraint on the timestep $\Delta t$, such that it has to be smaller than a critical value $\Delta t_{\rm{c}}$ for an explicit scheme (e.g. Leapfrog) to remain stable. Empirically, \citet{Monaghan1995} use the criterion:
\begin{equation}
\Delta t < \min \left(\frac{\rho}{K} \right),
\label{eq:crit_monaghan}
\end{equation}
which is essentially the minimum of the drag stopping time taken over all of the SPH particles.

\subsection{Von Neumann stability analysis}
  A more precise criterion can be derived by considering the stability of a simple explicit scheme such as the Forward Euler method. We consider the evolution of the drag terms over a single drag timetstep $\Delta t$, calculating the velocities at the timestep $n+1$ from the velocities at the timestep $n$. Considering only the time-discretisation of the equations, we have
\begin{eqnarray}
\frac{{\bf v}^{n+1}_{\rm g} - \mathbf{v}^{n}_{\rm g}}{\Delta t} & = & - \frac{K}{\hrhog} \left({\bf v}_{\rm g} - {\bf v}_{\rm d}\right), \label{eq:FE_scheme_1D1} \\
\frac{{\bf v}^{n+1}_{\rm d} - \mathbf{v}^{n}_{\rm d}}{\Delta t} & = & + \frac{K}{\hrhod} \left({\bf v}_{\rm g} - {\bf v}_{\rm d}\right),\label{eq:FE_scheme_1D2}
\end{eqnarray}
 We then perform a standard Von Neumann analysis, considering a perturbation of the velocity field with respect to equilibrium at the timestep $m$ corresponding to a monochromatic plane wave, i.e.
\begin{eqnarray}
{\bf v}^{m}_{\rm g} & = & {\bf V}^{m}_{\rm{g}} e^{ikx}, \label{eq:FE_perturb1}\\
{\bf v}^{m}_{\rm d} & = & {\bf V}^{m}_{\rm{d}} e^{ikx}, \label{eq:FE_perturb2}
\end{eqnarray}
where ${\bf v}^{m}_{\rm{g}}$ and ${\bf v}^{m}_{\rm{d}}$ are complex constants and $k$ is the wavenumber. Substituting Eqs.~\ref{eq:FE_perturb1} and \ref{eq:FE_perturb2} into Eqs.~\ref{eq:FE_scheme_1D1} and \ref{eq:FE_scheme_1D2} leads to the linear system
\begin{equation}
\left( 
\begin{array}{c}
{\bf V}_{\rm{g}} \\
{\bf V}_{\rm{d}}
\end{array}
\right)^{n+1} = \left(
\begin{array}{cc}
1 - \Delta t ~ \frac{K}{\hrhog} & \Delta t ~  \frac{K}{\hrhog}  \\
\Delta t ~ \frac{K}{\hrhod} & 1 - \Delta t ~  \frac{K}{\hrhod}
\end{array}
 \right)
 \left( 
\begin{array}{c}
{\bf V}_{\rm{g}} \\
{\bf V}_{\rm{d}}
\end{array}
\right)^{n}.
\label{eq:FE_syst}
\end{equation}
The two complex eigenvalues $\Lambda_{\pm,aj}$ of the matrix $\mathcal{M}$ are given by:
\begin{equation}
\Lambda_{\pm,ai} = 1 - \frac{\Delta t }{2} \left( \frac{K}{\hrhog} + \frac{K}{\hrhod}  \right) \pm \frac{\Delta t }{2} \left( \frac{K}{\hrhog} + \frac{K}{\hrhod}  \right) .
\label{eq:eigenvals}
\end{equation}
The condition for the numerical scheme to remain stable ($|\Lambda_{-}| < 1$) implies a minimum timestep given by
\begin{equation}
\Delta t < \Delta t_{\rm{c}} = \frac{\hrhog\hrhod}{K(\hrhog + \hrhod)} (\equiv t_{\rm s}).
\label{eq:crit_laibe}
\end{equation}
 We note that this expression differs slightly to the one suggested by \citet{Monaghan1995} (Eq.~\ref{eq:crit_monaghan}) as it involves the \emph{physical} drag stopping time $t_{\rm s}$ --- i.e. the typical time to damp the differential velocity between the gas and the dust fluids --- rather than $\frac{K}{\rho}$. The timestep of \citet{Monaghan1995} is thus correct in the limit where the density of one phase is negligible compared to the density of the other phase, but becomes erroneous in the case of two fluids having densities of the same order of magnitude. Note this would apply to grid codes also.
 
\subsection{SPH explicit timestep}
  The stability criterion for the full SPH system (and also for other explicit schemes) is expected to be similar to that derived for the continuum case (Eq.~\ref{eq:crit_laibe}). The main difference is that the drag coefficient $K$ is in general only defined on particle \emph{pairs} rather than individual particles. We thus take the minimum of Eq.~\ref{eq:crit_laibe} over a particle's neighbours, i.e.
\begin{equation}
\Delta t_{\rm{c}, a} = \min_{k} \left[ \frac{\hrhoga \hat{\rho}_{k} }{K_{ak}(\hrhoga + \hat{\rho}_{k} )} \right];  \hspace{0.5cm} \Delta t_{\rm{c}, i} = \min_{b} \left[ \frac{\hrhogb \hat{\rho}_{i} }{K_{bi}(\hrhogb + \hat{\rho}_{i})} \right]; 
\label{eq:crit_laibeSPH}
\end{equation}
for gas and dust particles, respectively. 

\subsection{Implicit timestepping}
For strong drag regimes, the timestep restriction imposed by Eq.~(\ref{eq:crit_laibeSPH}) becomes prohibitive, and an implicit timestepping algorithm is required, as proposed by \citet{Monaghan1997}. We use only explicit timestepping for the tests shown in this paper, with implicit timestepping methods discussed in detail in Paper~II.

\section{Numerical tests}
\label{sec:tests}
 Despite a number of codes having already been developed for simulating astrophysical gas-dust mixtures, none have been benchmarked against a wide range of test problems relevant to astrophysics. For example, while \citet{Monaghan1995}, \citet{Maddison2003} consider drag on a single dust particle in a box of gas (similar to our \textsc{dustybox} test below), no waves or shocks are considered. Similarly \citet{PM2006} benchmark their algorithm against a single dust-gas shock problem with only a qualitative solution. Other authors simply check that the timescale for settling in an accretion disc is roughly consistent \citep{BF2005} or provide no tests at all \citep{Rice04,Fromang2005,Fromang2006}. In the absence of known analytic solutions for simple problems, \citet{Johansen2007}, \citet{Miniati2010} and \citet{Bai2010} use the linear growth rates for the streaming instability \citep{Youdin2005} as a test problem, though this is already a complicated problem.

 In this paper we present a comprehensive suite of test problems designed to investigate all aspects of our algorithm relevant to astrophysics. These we refer to as \textsc{dustybox}, \textsc{dustywave}, \textsc{dustyshock}, \textsc{dustysedov} and \textsc{dustydisc}. Analytic solutions for the \textsc{dustybox} and \textsc{dustywave} problems have been derived in \citet{LP11a}, while the solution for \textsc{dustyshock} is known in the limit of high drag. The solutions for \textsc{dustysedov} and \textsc{dustydisc} are more qualitative but are important reference problems for astrophysical dust-gas mixtures. We consider simulation of the streaming instability to be of sufficient importance and complexity to be covered in detail in a separate paper. Note that all of the tests considered in this paper are performed, for simplicity, using a constant drag coefficient $K$. Realistic drag regimes are considered in Paper~II.

\begin{figure}
   \centering
   \includegraphics[angle=0, width=\columnwidth]{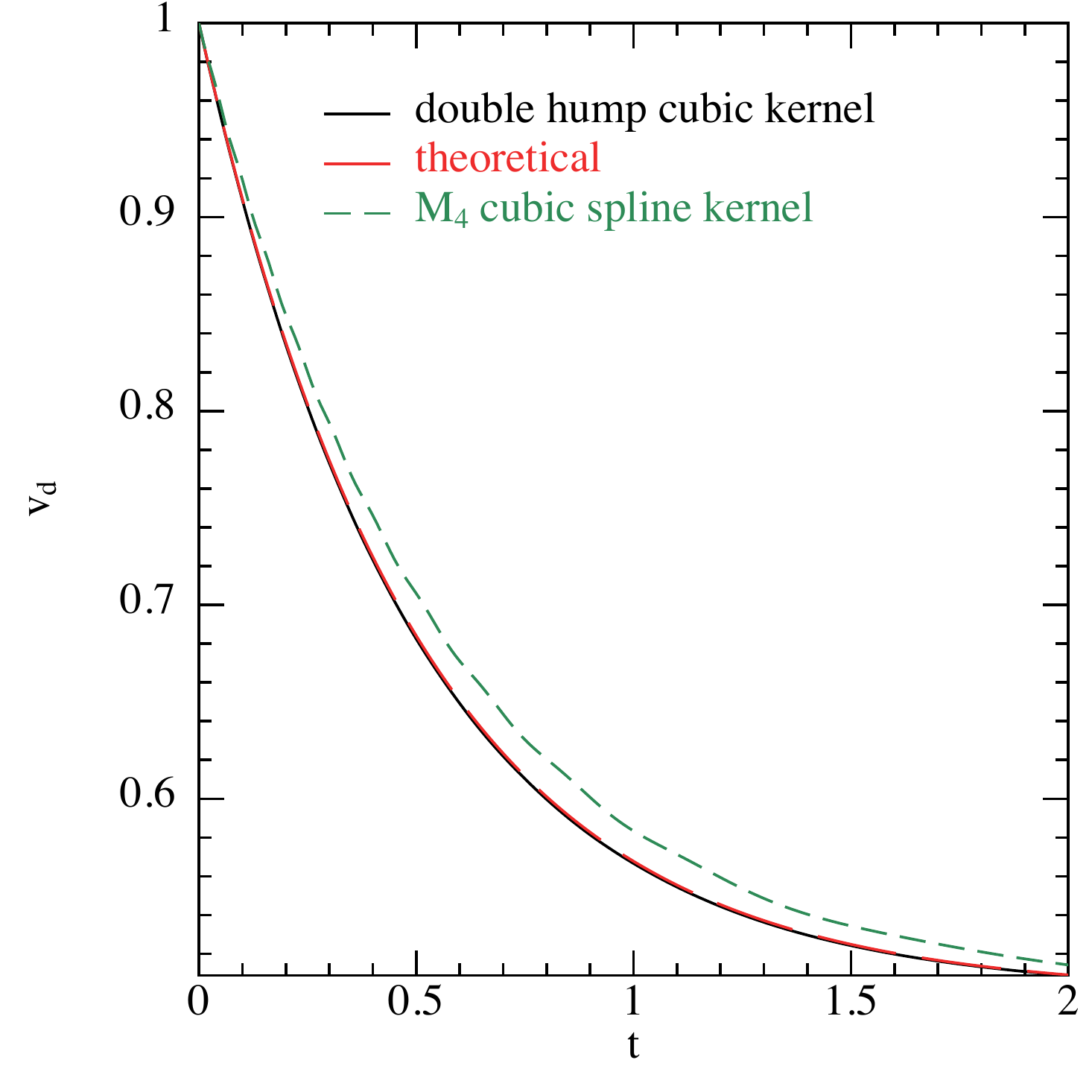} 
   \caption{Dust velocity as a function of time in the \textsc{dustybox} problem, using $2 \times 20^{3}$ particles, a dust-to-gas ratio of unity and a constant drag coefficient $K=1$. Use of the standard cubic spline kernel for the drag terms (short dashed/green line) results in errors of order $10\%$ in the velocities compared to the exact solution (long dashed/red line). Using a double-hump kernel (solid black) improves the accuracy --- for the same computational cost --- by a factor of several hundred, to $\lesssim 0.1\%$.}
   \label{fig:kernelcomp}
\end{figure}

\subsection{Code implementation}
 Implementation of the algorithm into a standard SPH code with explicit timestepping is relatively straightforward. The main changes are i) to store a particle type allowing setup and simulation of multiple fluids; ii) to compute the densities and smoothing lengths on each fluid as described in Sec.~\ref{sec:SPHdensities}; iii) compute the drag term between each fluid according to Eqs.~\ref{eq:SPH_drag_gas}--\ref{eq:SPH_drag_dust} and the heating term given by Eq.~\ref{eq:evol_SPH_energy_drag} and iv) (optional for astrophysics) to implement the modifications to the equations of motion due to the volume-filling fraction of the dust. We have implemented the two-fluid algorithm into both the $N-$dimensional \textsc{ndspmhd} test code \citep{Price2011} and into the parallel \textsc{phantom} code for 3D problems \citep{pf10,lp10}.

\subsection{\textsc{dustybox}: Two fluid drag in a periodic box}
\label{sec:dustybox}
 The \textsc{dustybox} problem presented by \citet{LP11a} involves two fluids in a periodic box moving with a differential velocity ($\Delta v _{0} =  v _{d,0} -  v _{g,0}$). It is similar to the test performed by \citet{Monaghan1995} showing the drag on a single dust grain in a box of gas, except that here we consider the dust as a fluid, meaning that the densities and smoothing lengths of both phases are computed self-consistently.

\subsubsection{\textsc{dustybox}: Setup}

 We setup the particles in a 3D periodic domain $x, y, z \in [0,1]$ such that the densities $\hrhog$ and $\hrhod$ and the gas pressure $\Pg$ are constant, and neglect the dust intrinsic volume by fixing the volume fraction $\theta = 1$. The box is filled by $20^{3}$ SPH gas particles set up on a regular cubic lattice and $20^{3}$ dust particles set up on a cubic lattice shifted by half of the lattice step in each direction. The gas sound speed, the gas and the dust densities are set to unity in code units and no artificial viscosity terms are applied. We give the fluids initial velocities $v_{\rm d} = 1$ and $v_{\rm g} = 0$.

During the simulation, we verified that both the total linear and angular momentum are exactly conserved as expected (Eqs.~\ref{eq:cons_mom_drag}--\ref{eq:cons_angmom_drag}). We have also verified that 1) the offset of the dust lattice with respect to the gas lattice and 2) the timestepping scheme do not affect the results.

\subsubsection{\textsc{dustybox}: Choice of drag kernel}

 Fig.~\ref{fig:kernelcomp} shows the dust velocity as a function of time in the \textsc{dustybox} test using $\hrhog = \hrhod = 1$ (i.e., a dust to gas ratio of unity) and $K=1$, with the exact solution from \citet{LP11a} shown by the long-dashed/red line. Using the cubic spline $M_{4}$ kernel (short-dashed/green line), the errors are of order 10\%. Since these errors are due to intrinsic bias in the kernel interpolation of the drag terms (Fig.~\ref{fig:kernelerrors}), they are independent of resolution, though can be improved -- at considerable cost -- by increasing the ratio of smoothing length to particle spacing (i.e., the neighbour number). By comparison, use of the double-hump cubic spline kernel gives errors $\lesssim 0.1\%$ (solid/black line) with no additional overhead in terms of cost.

\subsubsection{\textsc{dustybox}: Effect of drag coefficient and dust-to-gas ratio}

\begin{figure}
   \centering
   \includegraphics[angle=0, width=\columnwidth]{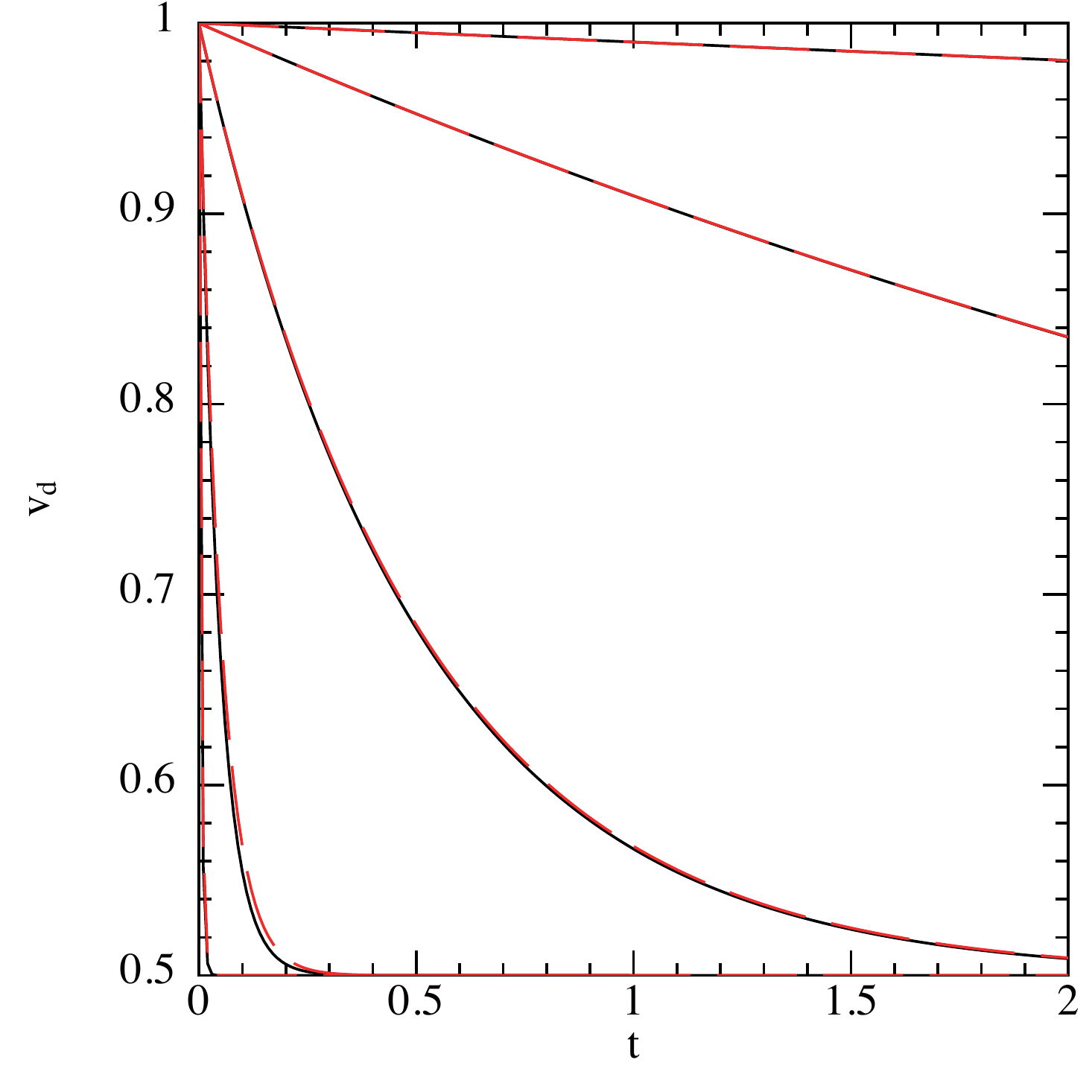} 
   \caption{As in Fig.~\ref{fig:kernelcomp} but using only the double hump cubic kernel with a range of drag coefficients $K = 0.01, 0.1, 1, 10, 100$ (top-to-bottom, solid/black lines), compared with the exact solution in each case given by the long-dashed/red lines.}
   \label{fig:Kbox}
\end{figure}

 Fig.~\ref{fig:Kbox} is identical to Fig.~\ref{fig:kernelcomp} but for a range of drag coefficients $K = 0.01, 0.1, 1, 10, 100$, compared to the exact solution in each case given by a solid/black line. Irrespective of the value of $K$, both gas and dust velocities relax to the barycentric velocity ($\vg = \vd = 0.5$) in a few stopping times $t_{\rm{s}} = (\hrhog \hrhod)/[K (\hrhog + \hrhod)]$. Using the double-hump cubic, an accuracy between 0.1 and 1\% is achieved in all cases (long dashed/red lines).

\begin{figure}
   \centering
   \includegraphics[angle=0, width=\columnwidth]{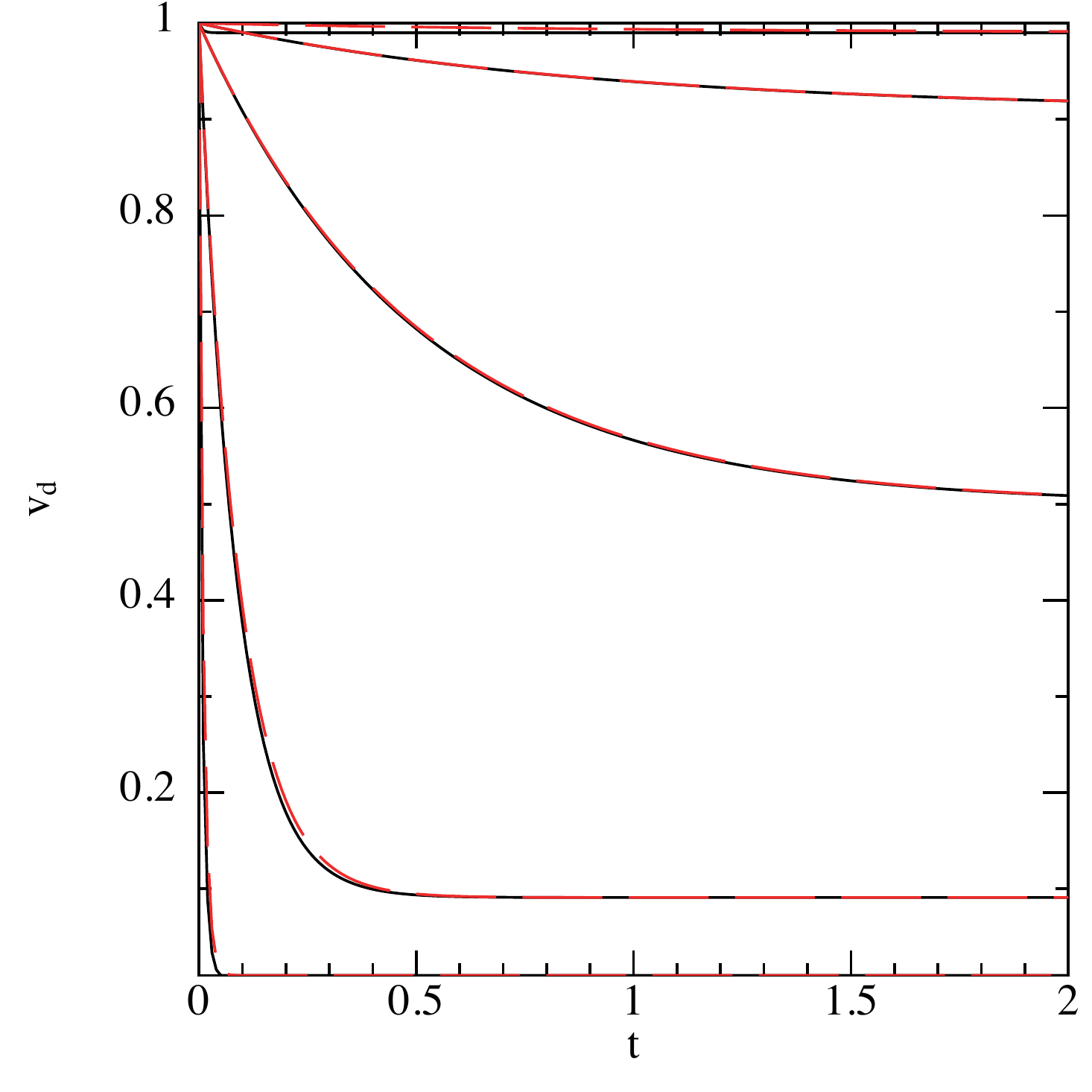} 
   \caption{As in Figs.~\ref{fig:kernelcomp} and \ref{fig:Kbox} but varying the dust-to-gas ratio $\hrhod / \hrhod = 0.01, 0.1, 1, 10, 100$ (top-to-bottom, solid/black lines) and a fixed drag coefficient $K=1$ using the double hump cubic kernel. Exact solutions for each case are given by the long-dashed/red lines.}
   \label{fig:dtgbox}
\end{figure}

 Fig.~\ref{fig:dtgbox} is similar, but varying the dust-to-gas ratio using $\hrhod / \hrhod = 0.01, 0.1, 1, 10, 100$ (achieved by varying $\hrhod$ with $\hrhog = 1$) and using $K=1$. This changes both the drag stopping time and the barycentric velocity towards which the system relaxes. Here again, an accuracy between 0.1 and 1\% is achieved in all cases.

\subsection{\textsc{dustywave}: Sound waves in a dust-gas mixture}
 The exact solution for linear waves propagating in a dust-gas mixture (\textsc{dustywave}) has been presented by \citet{LP11a}. We have performed a series of tests involving the propagation of a sound wave along the $x-$axis in both one and three dimensions in a periodic box, adopting the setup described in Table~2 of \citet{LP11a}. The \textsc{dustywave} problem is more complex than the \textsc{dustybox} problem as the motion of the mixture is driven by both the drag and the gas pressure.

Specifically,  \citet{LP11a} derive the dispersion relation:
\begin{equation}
\omega^{3} + iK \left(\frac{1}{\hrhog} + \frac{1}{\hrhod} \right)\omega^{2} -k^{2}c_{\rm s}^{2} \omega - iK \frac{k^{2}c_{\rm s}^{2}}{\hrhod} = 0, 
\label{eq:disprel}
\end{equation}
for solutions in the form $e^{i\left(kx - \omega t \right)}$. At high drag, Eq.~\ref{eq:disprel} can be expanded in a Taylor series, which to first order gives:
\begin{equation}
\omega = \pm k \tilde{c}_{\rm s} - i \frac{\hrhog \hrhod}{K \left(\hrhog + \hrhod \right)}k^{2} \cs^{2}\left(\frac{1 - A^{2}}{2} \right)
\label{eq:exp_omega}
\end{equation}
where the effective sound speed is defined according to
\begin{equation}
\tilde{c}_{\rm s} \equiv  c_{\rm s} A = c_{\rm s} \left(1 + \frac{\hrhod}{\hrhog}\right)^{-\frac12}.
\label{eq:csmod}
\end{equation}
The first term of Eq.~\ref{eq:exp_omega} gives the propagation of the centre of mass of the mixture at the effective sound speed $\tilde{c}_{\rm s}$. The second term corresponds to a corrective dissipative term since $A \in \left[0,1 \right]$.

\subsubsection{\textsc{dustywave}: Setup}
 The equilibrium state is characterised by the two phases at rest where the gas sound speed, and both gas and dust densities are set to unity in code units. In 1D this is achieved by placing equally spaced particles in the periodic domain $x \in [0,1]$. For the 3D simulations, the tests are run in a periodic box $x,y,z \in [0,1]$ with gas particles set up on a regular cubic lattice and dust particles set up on a cubic lattice shifted by half of the lattice step in each directions. As previously, no artificial viscosity is applied. We set the relative amplitude of the perturbation to $10^{-4}$ in both velocity and density in order to remain in the linear acoustic regime for which the solution in \citet{LP11a} is derived (we have verified that running the same simulations setting the relative amplitudes to $10^{-8}$ gives the same results). The density perturbation is applied to the particles as described in Appendix~B of \citet{pm04b}. We adopt an isothermal equation of state $P= c_{\rm s}^{2} \rho$ with $c_{\rm s} = 1$.

\begin{figure}
   \centering
   \includegraphics[angle=0, width=\columnwidth]{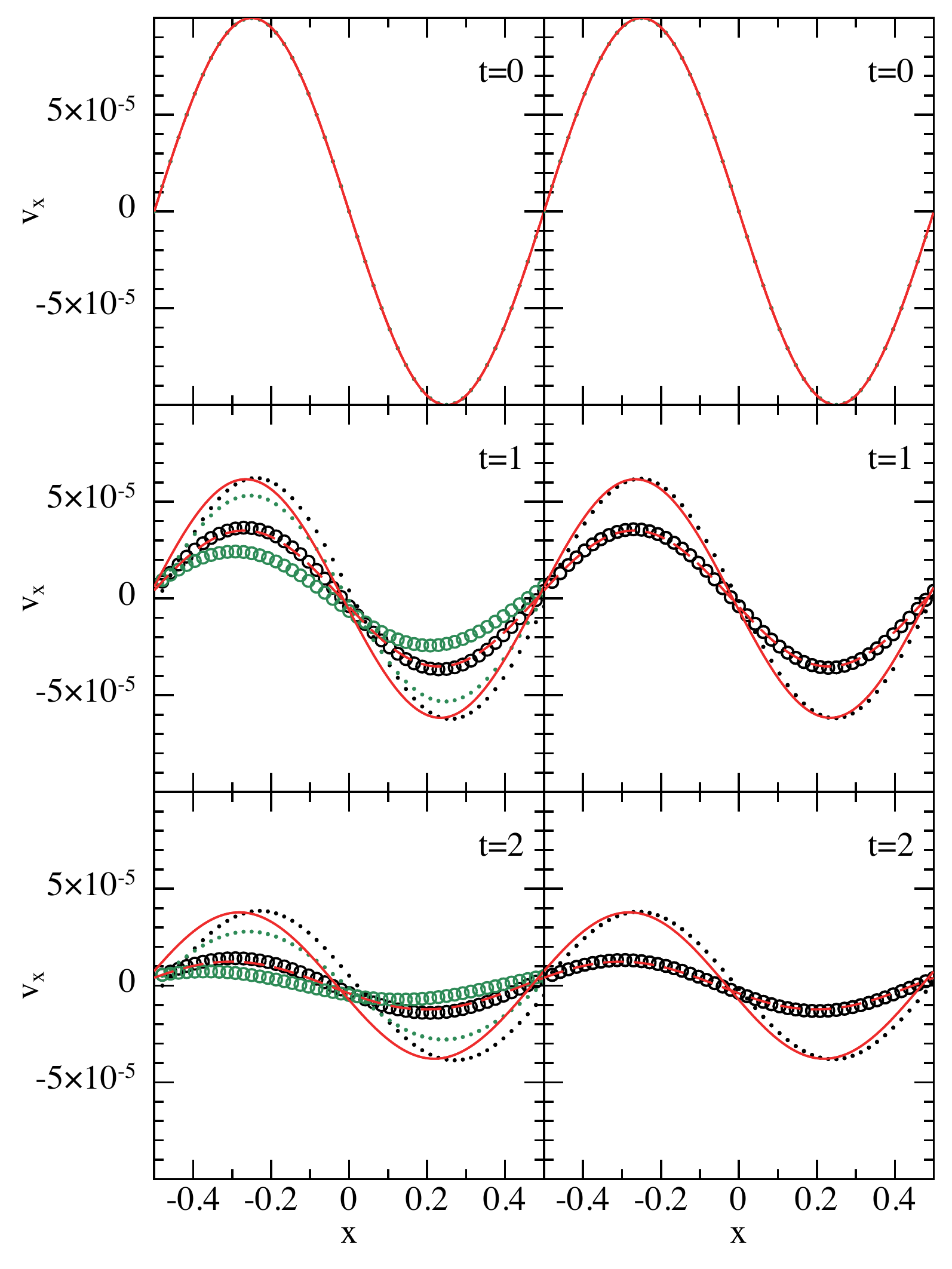} 
   \caption{Results of the \textsc{dustywave} test in 3D at $t=0$ (top row) and after 1 and 2 wave periods (middle and bottom rows) using $2 \times 32^{3}$ particles, $K=1$ and a dust-to-gas ratio of unity. The analytic solution is given by the solid/red (gas) and long-dashed/red (dust) lines. The standard cubic spline kernel (green points, left panel) performs poorly on this test. Using the double hump cubic kernel (black points, left panel) both the amplitude and frequency are correct but there remains a small phase error due to the kernel bias which can be corrected by using the smoother double-hump quintic kernel (black points, right panel).}
   \label{fig:wavekernel}
\end{figure}

\subsubsection{\textsc{dustywave}: Effect of the smoothing kernel}
 The results of the \textsc{dustywave} test in 3D using $2 \times 32^{3}$ particles with $K=1$ and a dust-to-gas ratio of unity is shown in Fig.~\ref{fig:wavekernel}, using the standard bell-shaped cubic (green points, lower amplitude) and double-hump cubic kernel (black points, correct amplitudes) for the drag (left panel) and the double-hump quintic kernel (c.f. Sec.~\ref{sec:dragkernel}) (right panel), at $t=0$ (top) and after 1 and 2 periods (middle and bottom panels). The numerical solutions (green and black markers) may be compared to the analytic solutions given by the solid/red (gas) and long-dashed/red (dust) lines. The amplitude and frequency of the solution are only correctly captured using double-hump kernels. With the double-hump cubic employed (left panel, black points) there remains a slight (few~\%) phase error in the numerical solution caused by the remaining kernel bias, independent of resolution. A similar error is found generically in multidimensional SPH simulations of linear waves (see e.g. Fig. 6 of \citealt{pm05}) and in standard SPH is improved by using a smoother kernel such as the quintic spline. Indeed, using the double-hump version of the quintic kernel for the drag terms and the standard quintic for the SPH terms (right panel) we find the phase error is smaller by a factor of $\sim5$.

 Longer multidimensional simulations of linear waves are more complicated in SPH because placement of the gas particles on regular lattices are unstable to low-amplitude transverse modes that cause the particles to rearrange towards a ``glass-like'' configuration \citep{morris96,morrisphd}. Furthermore, we find that with large drag coefficients we require extremely high resolutions to match the analytic solutions (see below), which becomes prohibitive in 3D. In one dimension however, the numerical stability of a sound wave is achieved simply by satisfying the courant condition (i.e. $\Delta t \leq 0.3 c_{\rm{s}}/h$) and the timestep constraint from the drag (Eq.~\ref{eq:crit_laibe}). We thus turn to 1D to investigate the full parameter range of the \textsc{dustywave} solution.

\begin{figure}
   \centering
   \includegraphics[angle=0, width=\columnwidth]{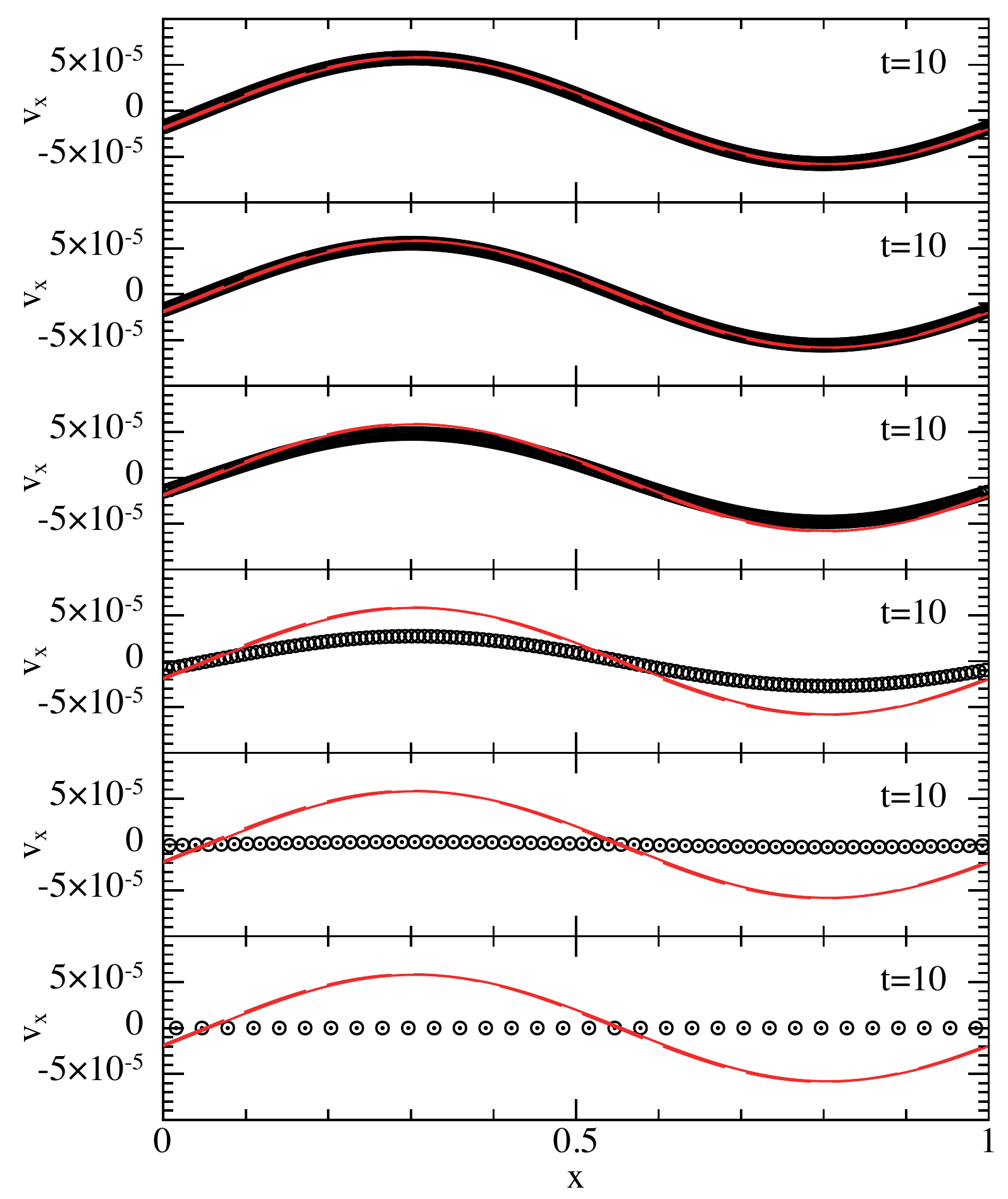} 
   \caption{Resolution study for the \textsc{dustywave} test in 1D using a high drag coefficient ($K=100$) and a dust-to-gas ratio of unity using 32, 64, 128, 256, 512 and 1024 particles from bottom to top. At large drag high resolution is required to resolve the small differential motions between the fluids and thus prevent over-damping of the numerical solution, corresponding to the criterion $h \lesssim c_{\rm s} t_{\rm s}$, here implying $\gtrsim 240$ particles. See also Fig.~\ref{fig:waveresekin}.}
   \label{fig:waveres}
\end{figure}

\begin{figure}
   \centering
   \includegraphics[angle=0, width=\columnwidth]{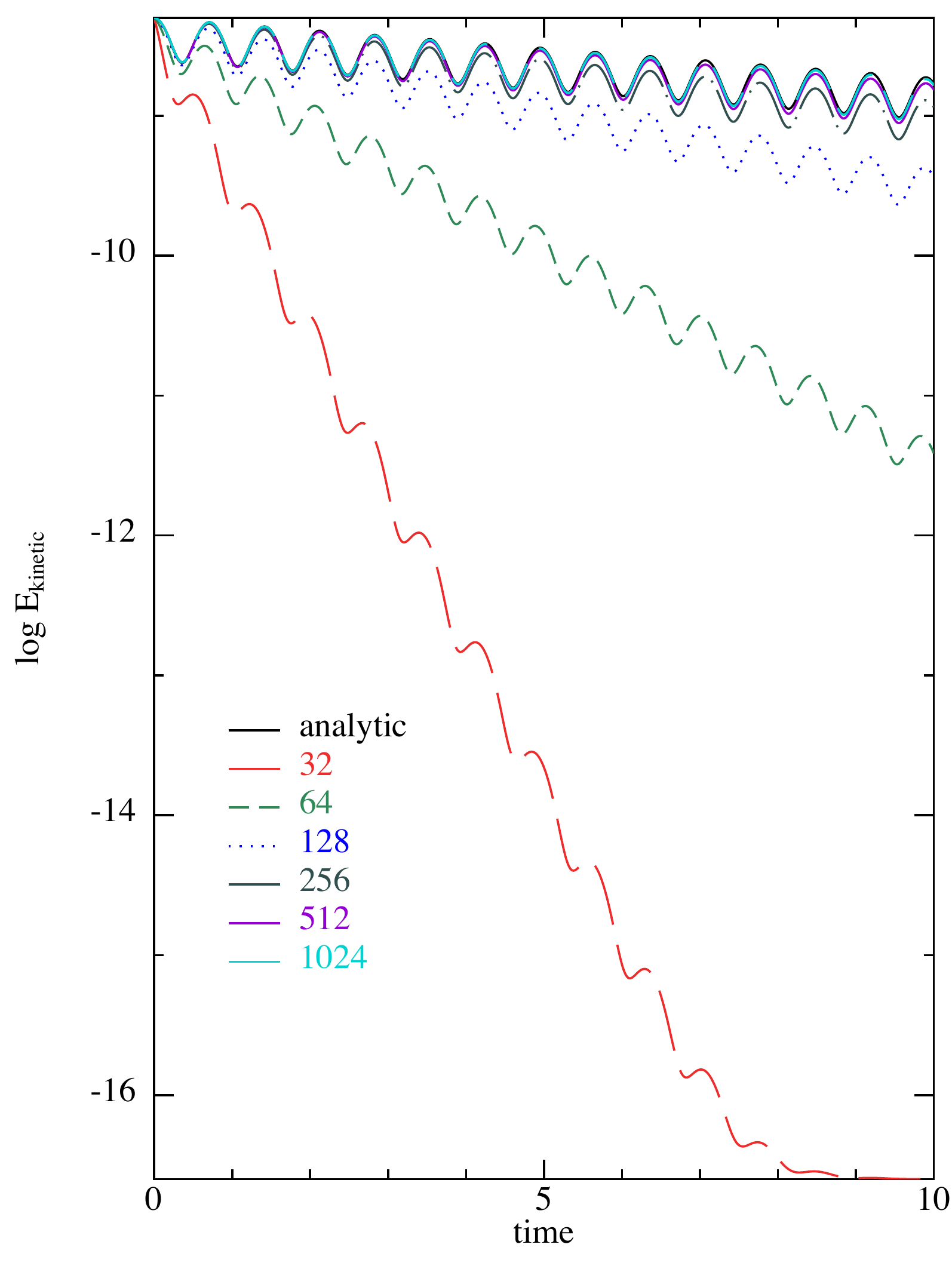} 
   \caption{As in Fig.~\ref{fig:waveres} but showing the kinetic energy as a function of time in the numerical solution at progressively increasing resolution, compared to the analytic solution given by the solid black line. The kinetic energy decay converges to the analytic solution at $\sim 256-512$ particles per wavelength, implying a demanding resolution criterion ($h \lesssim c_{\rm s} t_{\rm s}$) for high drag.}
   \label{fig:waveresekin}
\end{figure}

\begin{figure*}
   \centering
   \includegraphics[angle=0, width=\columnwidth]{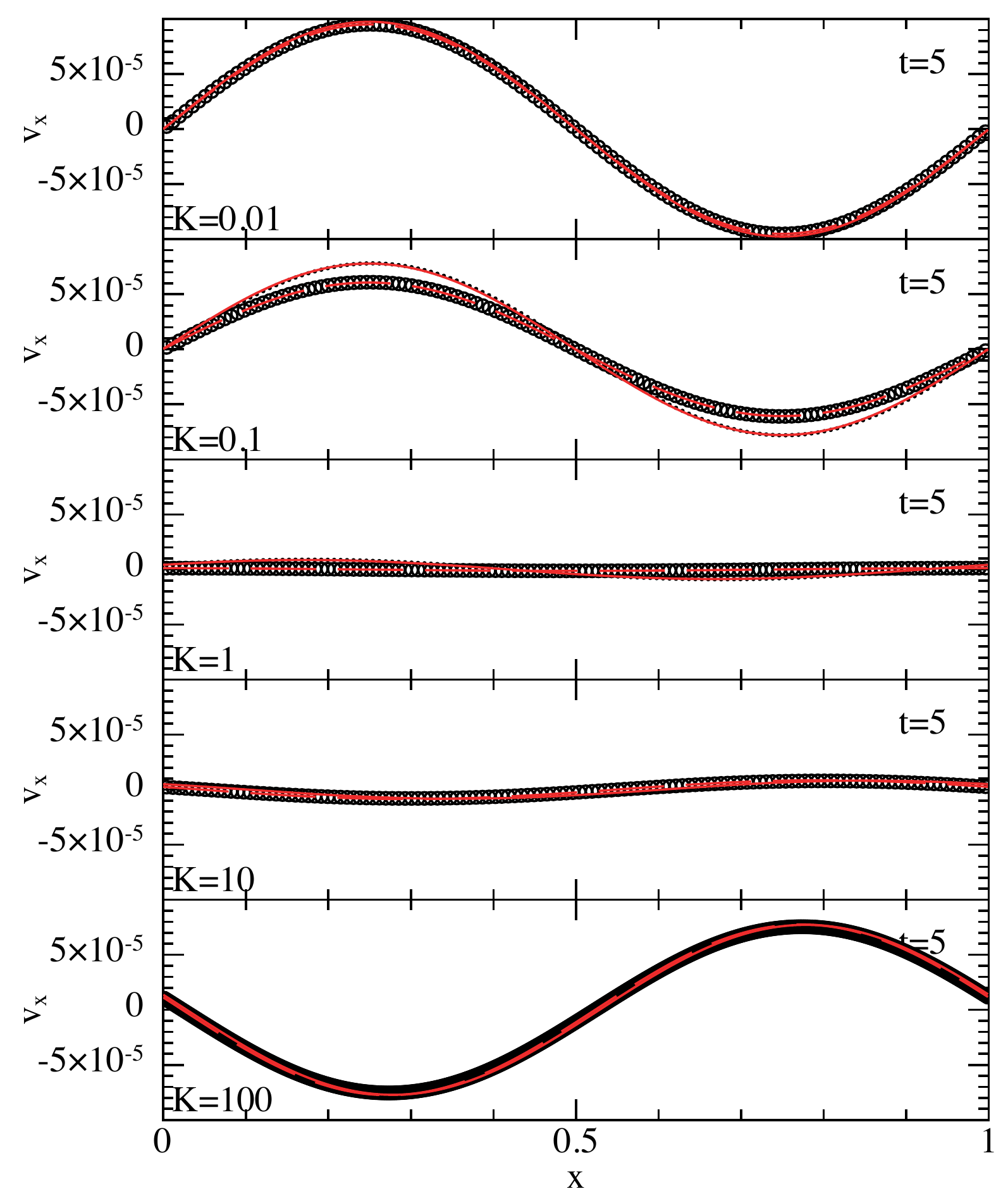}
   \hspace{0.5cm}
   \includegraphics[angle=0, width=\columnwidth]{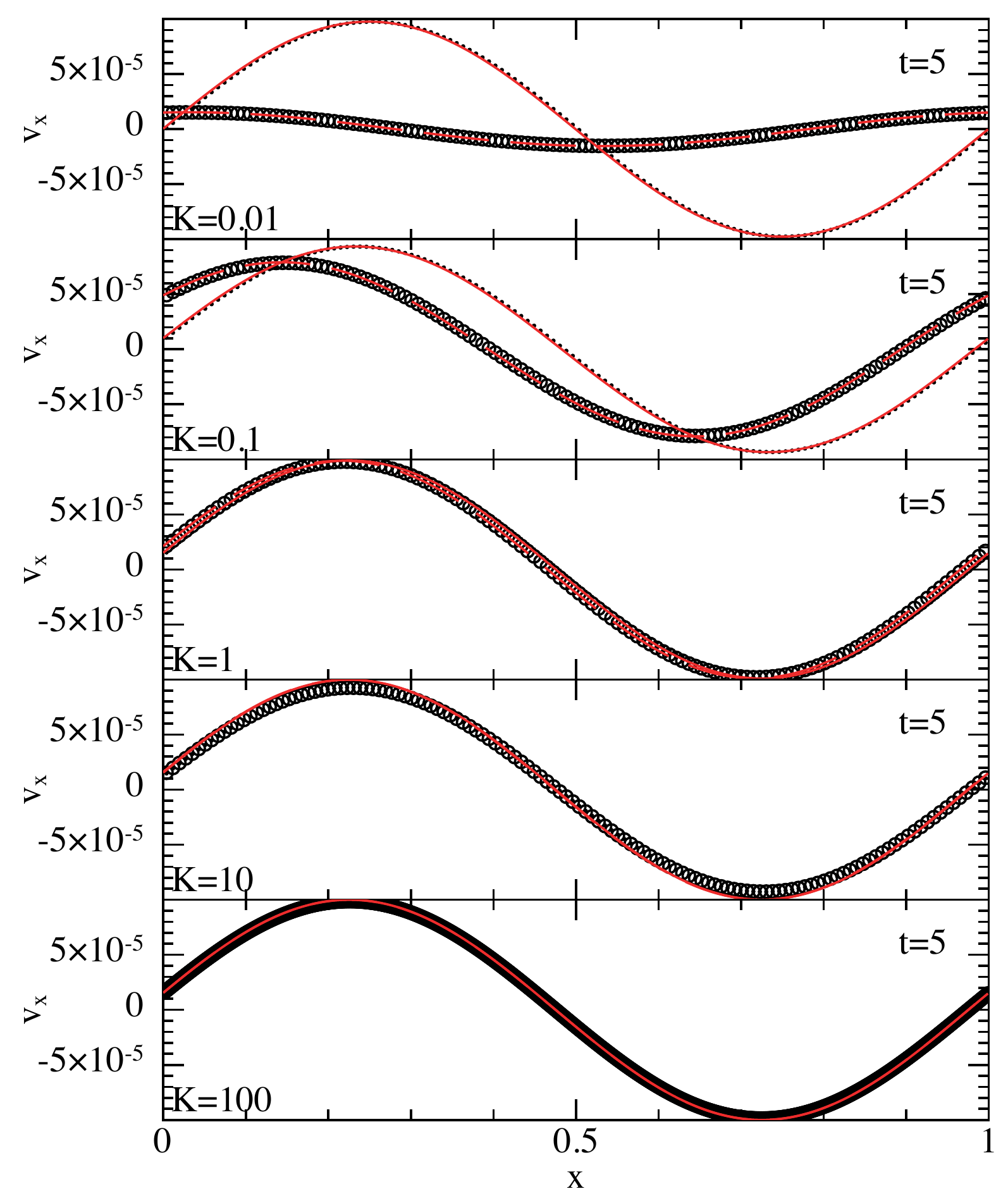} 
   \caption{Parameter study of the 1D \textsc{dustywave} problem, varying the drag coefficient from $0.01$ to $100$ (top to bottom) and using two different dust-to-gas ratios ($\hrhod/\hrhog = 1$, left figure and $\hrhod/\hrhog = 0.01$, right figure). Results are shown after 5 wave periods using 128 particles except for the $K=100$ case where $512$ particles are used to satisfy the resolution criterion $h \lesssim c_{\rm s} t_{\rm s}$. The double hump cubic kernel is used for the drag terms. The results obtained at these resolutions are indistinguishable from the analytic solutions (red solid [gas] and dashed [dust] lines) for both the gas (dots) and the dust (circle) particles.}
   \label{fig:waveparam}
\end{figure*}

\subsubsection{\textsc{dustywave}: Resolution requirements at high drag}
 Fig.~\ref{fig:waveres} shows the velocity profiles after 10 periods in the 1D \textsc{dustywave} problem for a large drag coefficient ($K=100$) and a dust-to-gas ratio of unity, with numerical resolution as indicated. At low resolutions ($\lesssim 256$ particles per wavelength) and high drag, the amplitude of the wave in the numerical simulations (black solid [gas] and open [dust] circles, on top of each other) is severely overdamped compared to the analytic solution (red solid and long-dashed lines, also on top of each other). This is further illustrated in Fig.~\ref{fig:waveresekin} which shows the kinetic energy as a function of time for simulations at different resolutions, compared to the analytic solution given by the solid red line.

 Figs.~\ref{fig:waveres} and \ref{fig:waveresekin} illustrate a key difficulty that arises when considering high drag coefficients, i.e., where the drag stopping time $t_{\rm{s}}$ defined in Eq.~\ref{eq:crit_laibe} is much smaller than the period $T$ of the wave. In this case, the drag term efficiently damps the initial differential velocity between the gas and the dust in a few $t_{\rm{s}}$. However, as the pressure continues to drive the propagation of the wave in the gas, a small residual de-phasing of order $\sim c_{\rm s} t_{\rm s}$ occurs, which is simply the distance travelled by the gas before it is damped by the dust. This de-phasing induces a small differential velocity which in turn be damped by the drag. This small differential effect dissipates the kinetic energy on a timescale $\sim t_{\rm{s}}$.

 The spatial de-phasing between the gas and the dust represents the smallest length of the problem that must be resolved numerically in order to capture the physics of the process. If the spatial de-phasing between the gas and the dust is under-resolved, the differential velocity between the gas and the dust is artificially larger than the theoretical one, leading to a non-physical over-dissipation of the kinetic energy of the system, as observed in Figs.~\ref{fig:waveres} and \ref{fig:waveresekin}.
 
  We thus propose a resolution criterion for resolving the differential drag of the form
\begin{equation}
\Delta \lesssim c_{\rm s} t_{\rm s},
\label{eq:dragresgene}
\end{equation}  
where $\Delta$ is the resolution length. For SPH, this becomes
\begin{equation}
h \lesssim c_{\rm s} t_{\rm s}.
\label{eq:dragres}
\end{equation}
 For $K=100$ and $c_{\rm s} = \hrhog = \hrhod = 1$ in code units this implies $h < 0.02$, i.e. a minimum of $\sim 240$ particles (assuming $\eta = 1.2$ in Eqs.~\ref{eq:h_density_gas} and \ref{eq:h_density_dust}), which is consistent with Figs.~\ref{fig:waveres} and \ref{fig:waveresekin}.

Simulating dust-gas interactions at high drag therefore requires a high spatial resolution in order to accurately resolve the propagation without over-dissipating the energy of the system. This can lead to a prohibitive computational cost, somewhat counterintuitively since the drag simply tends to make the dust stick to the gas. Most importantly, this requirement is not unique to SPH and is a critical issue for any numerical method. Indeed, \citet{Bai2010} find similarly high resolution requirements at short stopping times in their simulations of the streaming instability.

\subsubsection{\textsc{dustywave}: Parameter study}

 Fig.~\ref{fig:waveparam} shows the results of 1D simulations of the \textsc{dustywave} problem for 5 drag coefficients (from $K = 10^{-2}$ to $K = 10^{2}$) and two different dust to gas ratio relevant for astrophysical systems ($1$ and $0.01$ for left and right figures, respectively), showing the velocities after five periods compared with the analytic solutions in each case. The simulations employ 128 particles except for the $K=100$ case where $512$ particles have been used in order to satisfy the criterion (\ref{eq:dragres}). For this set of parameters, our method provides results with an excellent accuracy (better than one per cent) on the frequencies, the amplitudes and the phases of both the gas and the dust velocities (and consequently the energy of the system).

For equal dust to gas ratios ($\hrhod/\hrhog = 1$, left figure), both phases are equally affected by the drag. At low drag ($K = 0.01$, top panel of left figure), the damping is not efficient enough for gas or the dust to be damped as the stopping time is $\sim$ hundreds of periods. At intermediate drag ($K = 1$, middle panel of left figure), the damping is the most efficient for the two phases. At large drag regimes ($K = 100$), the damping of the differential velocity occurs quickly, but the dust density is large enough to distort the gas propagation: the wave is de-phased by a half-period compared to the gas-only solution.

 With more typical astrophysical dust to gas ratios ($\hrhod/\hrhog = 0.01$, right figure), the gas remains essentially unaffected by the dust. It thus propagates almost freely in the box at a velocity close to the sound speed. By contrast, the dust phase is strongly affected by the drag as shown by the $K = 0.01$ case (top panel, right figure), where the damping time for the dust phase is $\sim$ one period. The differential velocity between the two phases becomes more and more efficiently damped as the drag coefficient increases (right panel, from top to bottom), making the dust phase stick to the gas.

\subsection{\textsc{dustyshock}: shock tube in a dust-gas mixture}
Propagation of a shock in a two-fluid dust and gas mixture (the \textsc{dustyshock} problem hereafter) has been studied both analytically (see e.g. \citealt{Rudinger1964}) and numerically (see e.g. \citealt{Miura1982,Saito2003}), using grid based methods. The \textsc{dustyshock} occurs in two stages: a transient stage (for which no analytic solution is known and therefore studied numerically) followed by a stationary stage which consists of the solution for a pure gas solution propagating at a modified $\gamma$ and the modified sound speed (Eq.~\ref{eq:csmod}, see also \citealt{Miura1982}). In an astrophysical context, simulations of a dusty shock were used by \citet{PM2006} to test their Godunov-type scheme using a Roe Solver developed to simulate astrophysical dust and gas mixtures.

The hypothesis for the dust phase in these seminal studies are essentially the same as the ones used in this paper. However, unnecessary additional complications arise from their choice of the Stokes drag regime (a function of the local Reynolds number for the particles), the addition of a heat transfer term (depending on the dust conductivity and the Nusselt number of the system) and a temperature-dependent gas viscosity. For the purposes of benchmarking of our numerical scheme, we instead simulate a simplified problem: using a linear drag regime with constant drag term $K$, no heat transfer between the phases and no viscosity other than the standard shock-capturing terms used in SPH. While the evolution during the transient stage may be different from those considered in previous studies, the solution during the stationary stage remains unchanged.

\subsubsection{\textsc{dustyshock}: setup}
We setup the \textsc{dustyshock} problem as a two fluid version of the standard \citet{sod78} problem. Equal mass particles are placed in the 1D domain $x \in [-0.5, 0.5]$, where for $x<0$ we use $\rhog = \rhod = 1$, $v_{\rm{g}} = v_{\rm{d}} = 0$ and $P_{\rm{g}} = 1$, while for $x>0$ $\rhog = \rhod = 0.125$, $v_{\rm{g}} = v_{\rm{d}} = 0$ and $P_{\rm{g}} = 0.1$. We use an ideal gas equation of state $P = (\gamma - 1)\rho u$ with $\gamma = 5/3$. The density jump means that for SPH the resolution is 8 times higher to the left of the shock than to the right. We adopt the same initial resolution in both the gas and the dust. This differs slightly from the setup used by \citet{Miura1982} and \citet{Saito2003} where the dust is only placed in the right half of the box. Standard artificial viscosity and conductivity terms are employed for shock-capturing in SPH as described in \citet{Price2011} with constant coefficients $\alpha_{\rm SPH} = 1$, $\beta_{\rm SPH} = 2$ and $\alpha_{\rm u} = 1$.

\subsubsection{\textsc{dustyshock}: transient evolution}

\begin{figure}
   \centering
   \includegraphics[angle=0, width=\columnwidth]{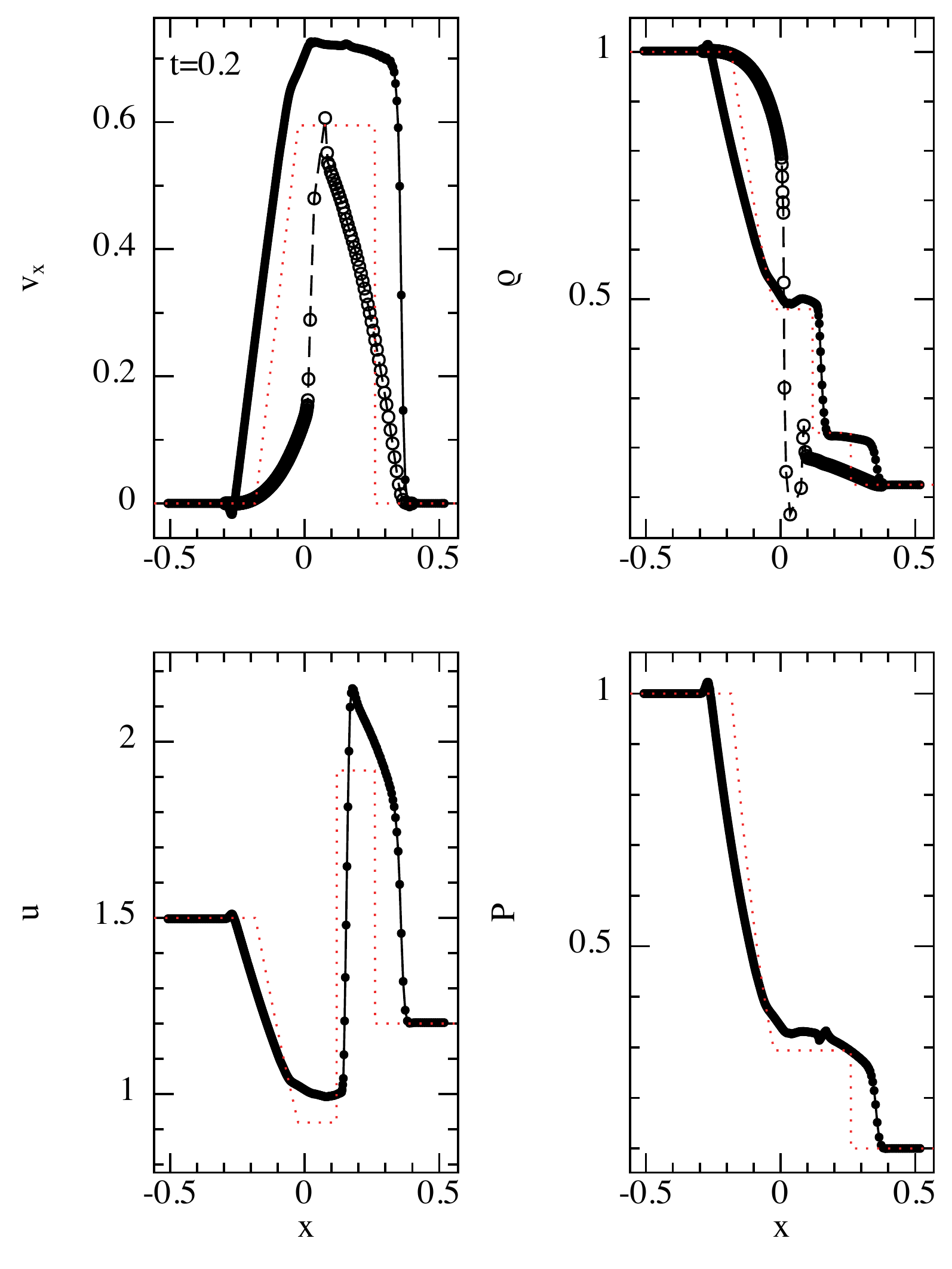} 
   \caption{Results of the \textsc{dustyshock} problem with a moderate drag coefficient $K=1$ and a dust-to-gas ratio of unity. This shows the \textsc{dustyshock} solution during the transient stage where the analytic solution is not known (the solution for the later stationary stage is shown by the dotted red line for comparison). Results are similar to those obtained in previous studies \citep{Miura1982,Saito2003}. Top panels show velocity and density in both gas (solid points) and dust (open circles), while bottom panels show thermal energy and pressure in the gas. Initial particle spacing to the left of the shock in both fluids is $\Delta x = 0.001$ while to the right it is $\Delta x = 0.008$, giving $569$ equal mass particles in each phase.}
   \label{fig:shocktrans}
\end{figure}

\begin{figure*}
   \centering
   \includegraphics[angle=0, width=\columnwidth]{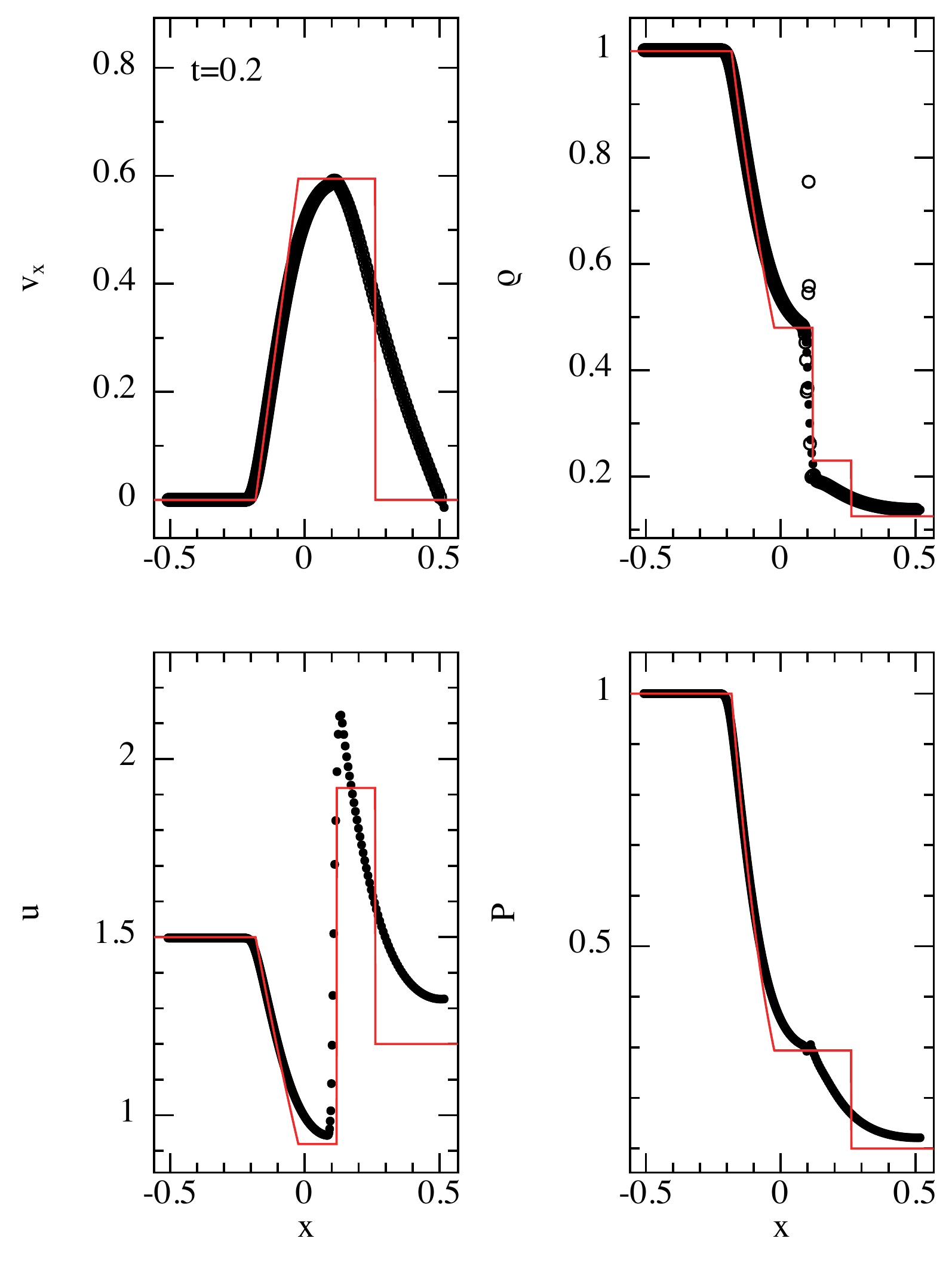} 
   \hspace{0.5cm}
   \includegraphics[angle=0, width=\columnwidth]{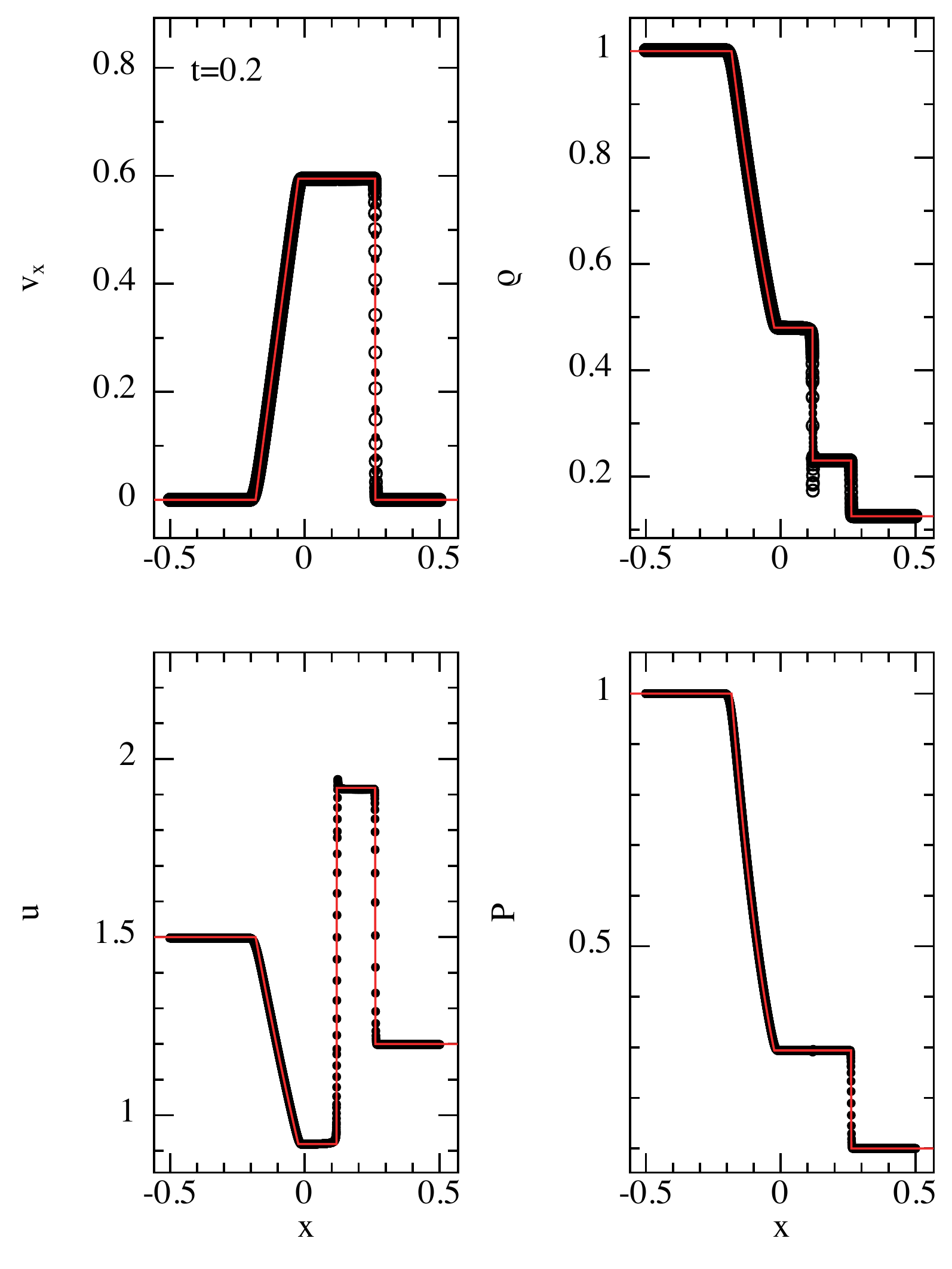} 
   \caption{Results of the \textsc{dustyshock} problem with a high drag coefficient ($K=1000$) and a dust-to-gas ratio of unity, thus being in the stationary phase where the analytic solution is known (solid/red lines). At low resolution (left figure, same resolution as Fig.~\ref{fig:shocktrans}) the results are incorrect due to the failure to satisfy the resolution criterion at high drag (Eq.~\ref{eq:dragres}). With this criterion satisfied (right panel, using $2 \times 11255$ particles) the numerical solution faithfully reproduces the analytic result. Thus, as in the \textsc{dustywave} test, extremely high resolution is required to obtain the correct solution at high drag.}
   \label{fig:shockstatio_high}
\end{figure*}

Fig.~\ref{fig:shocktrans} shows the results of a simulation using $2 \times 569$ particles (i.e. a particle spacing of $\Delta x = 0.001$ for $x < 0$) with a moderate drag coefficient ($K = 1$) and a dust-to-gas ratio of unity, showing velocity and density for both the gas and dust (top panels) and the thermal energy and pressure in the gas (lower panels). With this choice of drag coefficient the system remains in the transient regime at the time shown (since $t < t_{\rm s}$). It should be noted that while there is no known analytic solution for this stage of the problem, the shock profile we obtain is similar to those found previously (see e.g. \citealt{Miura1982,Saito2003}). Initially as the shock propagates in the mixture, the dust (initially at rest) dissipates the momentum and kinetic energy from the gas, lowering the propagation velocity compared to the ideal (gas only) case (dotted red line). The dust density ramps up roughly linearly behind the shock, reaching a density near the contact discontinuity roughly twice the unshocked dust density. \citet{Saito2003,Miura1982} and \citet{PM2006} also found a similar behaviour, also with a factor of 2 increase in the dust density behind the shock. The gas-dust interaction to the left of the contact discontinuity and in the rarefaction wave has not been previously studied since the above authors place the dust only downwards of the shock front. We find that the dust density decreases to near zero upstream of the contact discontinuity, increasing sharply at the head of the rarefaction wave, transitioning smoothly through the rarefaction wave to match the undisturbed value. We have checked that increasing the resolution further does not change the solution significantly for this choice of drag parameters.

\subsubsection{\textsc{dustyshock}: stationary regime}

 Fig.~\ref{fig:shockstatio_high} shows the results of simulations with a high drag coefficient ($K=1000$) and a dust-to-gas ratio of unity. In this case, since $t > t_{\rm s}$, the mixture quickly reaches the stationary regime. We are thus able to compare the SPH results to the analytic solution given by the solid red line (this corresponds to the standard hydrodynamic shock solution with modified sound speeds given by Eq.~\ref{eq:csmod}). The left figure shows the results at low resolution ($2 \times 569$ particles, as in Fig.~\ref{fig:shocktrans}, while the right figure shows the results at $20 \times$ higher resolution ($2 \times 11255$ particles). As in the \textsc{dustywave} test, we find that at high drag an extremely high resolution is required to obtain the correct solution, consistent with our resolution criterion derived above (Eq.~\ref{eq:dragres}). If this criterion is not satisfied the numerical shock solution is strongly inaccurate (left panel).

\begin{figure*}
   \centering
   \includegraphics[angle=0, width=0.72\textwidth]{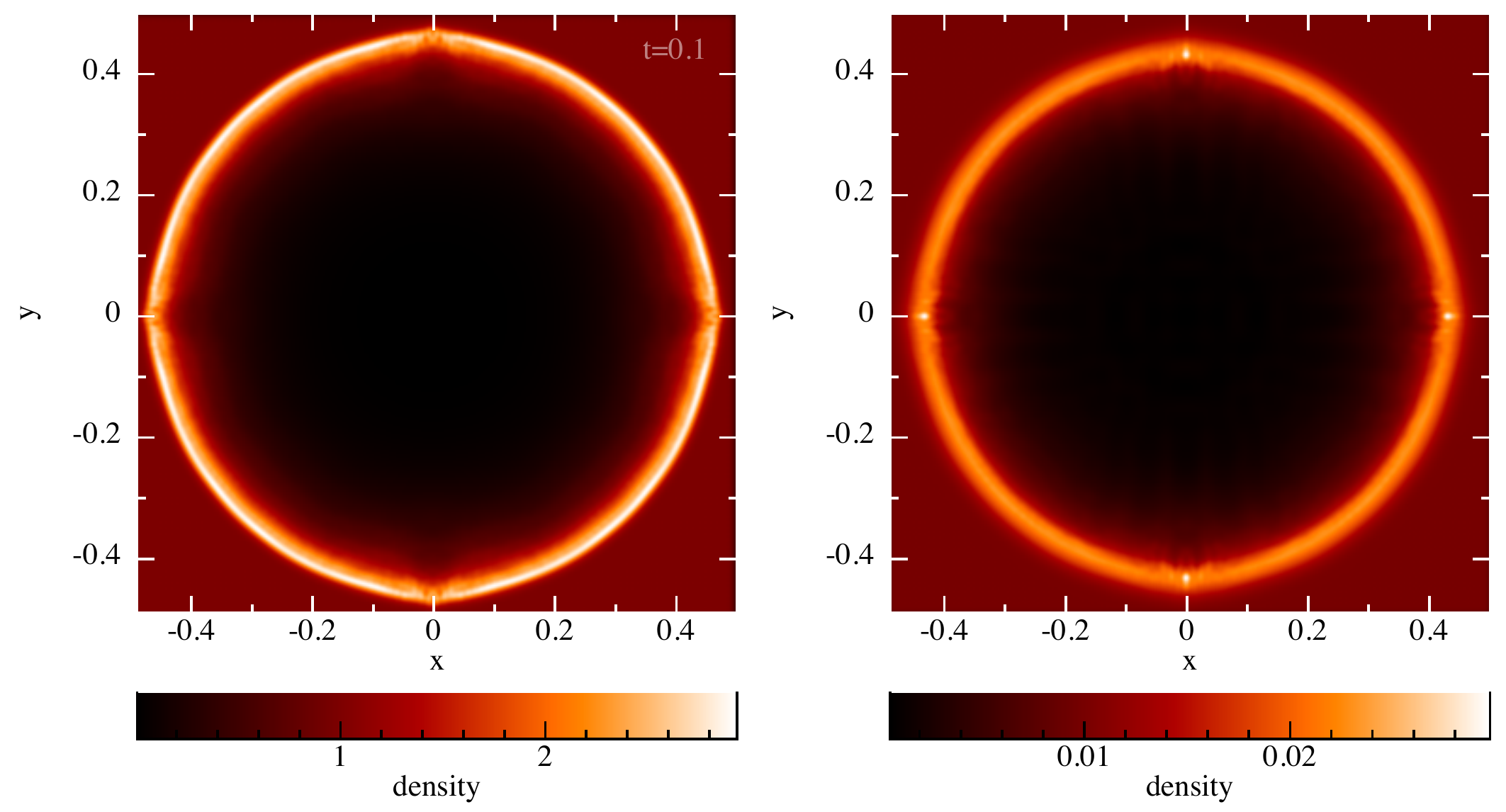} 
   \caption{Cross-section slice showing density in the midplane in the 3D \textsc{dustysedov} problem, for both the gas (left panel) and the dust (right panel) at $t=0.1$. A dust-to-gas ratio of $0.01$ and a drag coefficient of $K=1$ have been used with $100^{3}$ SPH particles in each phase. Note the slight difference in the blast radius between the dust and the gas, consistent with the response time ($t_{\rm s}$) of the dust to the gas drag.}
   \label{fig:render_K}
\end{figure*}

\begin{figure}
   \centering
   \includegraphics[angle=0, width=0.49\columnwidth]{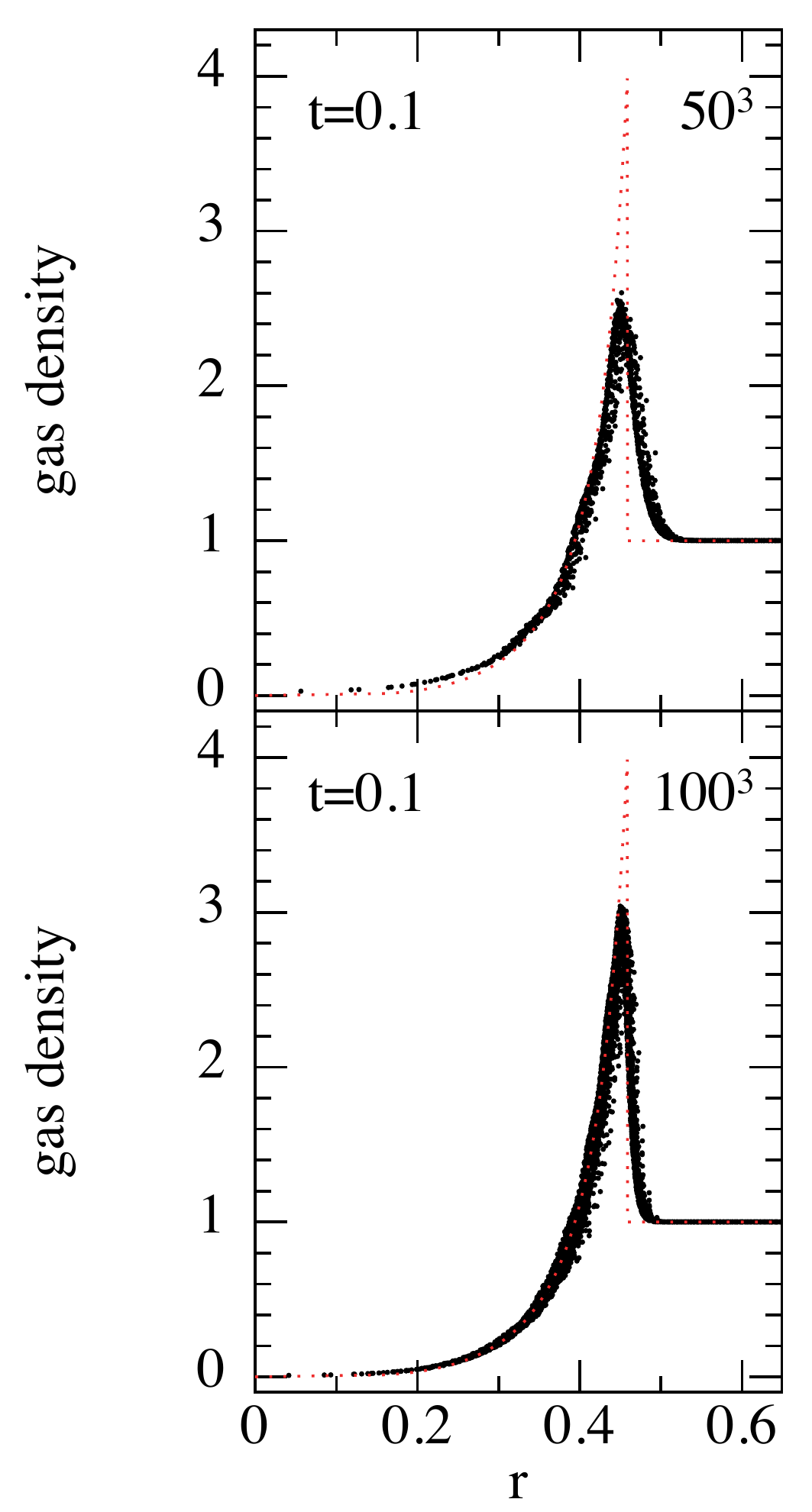} 
   \includegraphics[angle=0, width=0.49\columnwidth]{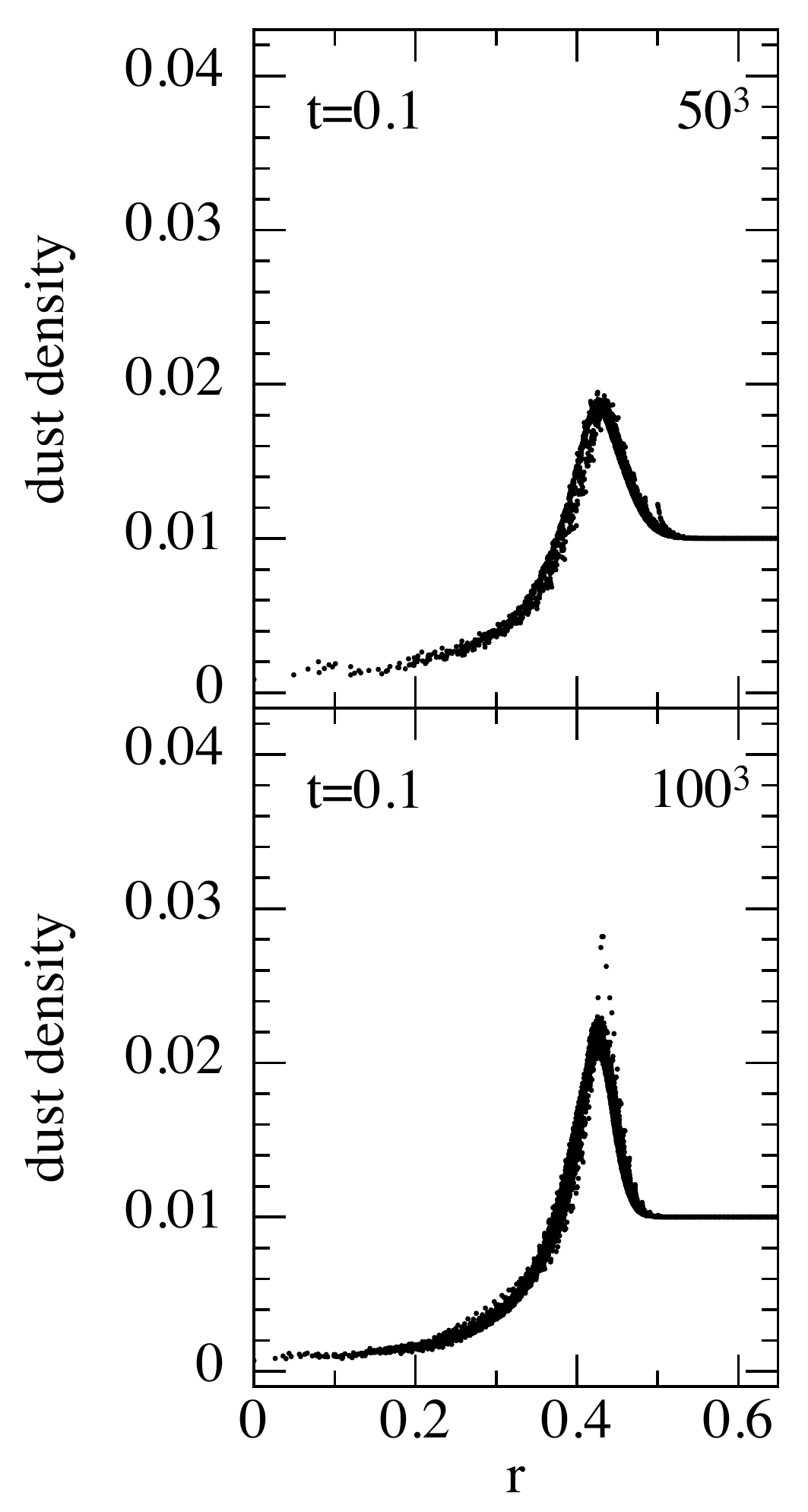} 
   \caption{Results of the 3D \textsc{dustysedov} test, showing the density in the gas (left figure) and dust (right figure) from a Sedov blast wave propagating in an astrophysical ($1\%$ dust-to-gas ratio) mixture of gas and dust with a constant drag coefficient $K = 1$. The low dust-to-gas ratio means that the gas is only weakly affected by the drag from the dust, and is thus close to the self-similar Sedov solution (dotted/red line). The dust density is affected by the propagation of the blast, resulting in an overdensity that closely mirrors the gas overdensity. Results are shown using $50^{3}$ (top panels) and $100^{3}$ particles (bottom panels).}
   \label{fig:sedov_K_density}
\end{figure}

\subsection{\textsc{dustysedov}: Sedov blast waves in a dust-gas mixture}
 The \textsc{dustysedov} test concerns the propagation of a Sedov blast wave in a dust-gas mixture. Although the self-similar Sedov solution is known for the propagation of a blast wave in a gas phase, the solution for a two fluid dust-gas mixture is unknown (though at high drag as previously it may be expected that the solution should revert to the gas-only solution using the modified sound speed). We do not attempt to simulate this problem at high drag as it would involve a prohibitive computational expense, instead adopting an ``astrophysical'' dust-to-gas ratio of $1\%$ and a moderate drag coefficient such that the presence of dust represents only a small perturbation to the gas evolution. Results of purely hydrodynamic SPH solutions for this test can be found e.g. in \citet{rp07} and \citet{sh02}.

\subsubsection{\textsc{dustysedov}: Setup}
 We setup the \textsc{dustysedov} problem in a 3D periodic box (the boundary conditions are irrelevant for the times shown) at two different resolutions, filling the box $x,y,z \in [-0.5, 0.5]$ by $50^{3}$ and $100^{3}$ SPH particles for both the gas and the dust. Gas particles are set up on a regular cubic lattice, with the dust particles also on a cubic lattice but shifted by half of the lattice step in each direction. We use $\alpha_{\rm SPH} = 1$ and  $\beta_{\rm SPH} = 2$ in the artificial viscosity terms, and $\alpha_{\rm u} = 1$ in the artificial conductivity term. An ideal gas equation of state $P = (\gamma -1)\rho u$ is adopted with $\gamma = 5/3$.
 
 In the self-similar Sedov solution, the thermal energy of the gas is initially concentrated at $r=0$. In the SPH simulation we distribute the internal energy of the gas over the particles located inside a radius $r < r_{\rm{b}}$ where $r_{\rm b}$ is set to 2h (i.e., the radius of the smoothing kernel). In code units the total blast energy is $E=1$, with $\hrhog = 1$ and $\hrhod = 0.01$. For $r>r_{\rm{b}}$, the gas sound speed is set to be $2 \times 10^{-5}$ in code units. The dust-to-gas ratio is set to $0.01$ to be consistent with the value measured for the interstellar medium. The drag coefficient is set to $K = 1$. Translated to physical units, assuming a box size of $1$ pc, an ambient sound speed of $2 \times 10^{4}$ cm/s and a gas density of $\rho_{0} = 6 \times 10^{-23}$ g/cm$^{3}$ the energy of the blast is $2 \times 10^{51}$ erg and time is measured in units of $100$ years, roughly corresponding to a supernova blast wave propagating into the interstellar medium. Obviously in a real supernova the temperature inside the blast would be much higher than the sublimation temperature of the dust, meaning that it would be quickly evaporated, so the \textsc{dustysedov} test is mainly useful as a benchmarking problem.

\subsubsection{\textsc{dustysedov}: Results}

 Fig.~\ref{fig:sedov_K_density} shows the densities of both the gas (left figure) and the dust phase (right figure) at $t=0.1$ using $50^{3}$ (top) and $100^{3}$ (bottom) particles for both fluids. Fig.~\ref{fig:render_K} shows a cross section of the density in the midplane in the high resolution ($100^{3}$) simulation, showing the gas (left) and dust (right). As the gas and dust densities are $1$ and $0.01$, respectively and the drag coefficient is $K=1$ in code units, the stopping time is $t_{\rm{s}} = 0.01$, which represents $10\%$ of the time required for the blast to fill the box. The response of the dust to the forcing by the gas drag is therefore of order $10\%$. Consequently, an overdensity in the dust phase forms due to the passage of the overdensity in the gas. The overdensities in the gas and the dust phases are dephased slightly (seen by comparing the position of the peak densities in the gas and dust in Fig.~\ref{fig:sedov_K_density}), consistent with the finite time ($t_{\rm{s}}$) required for the dust to respond to the gas forcing.
 
 \begin{figure*}
  \centering
   \includegraphics[angle=0, width=\textwidth]{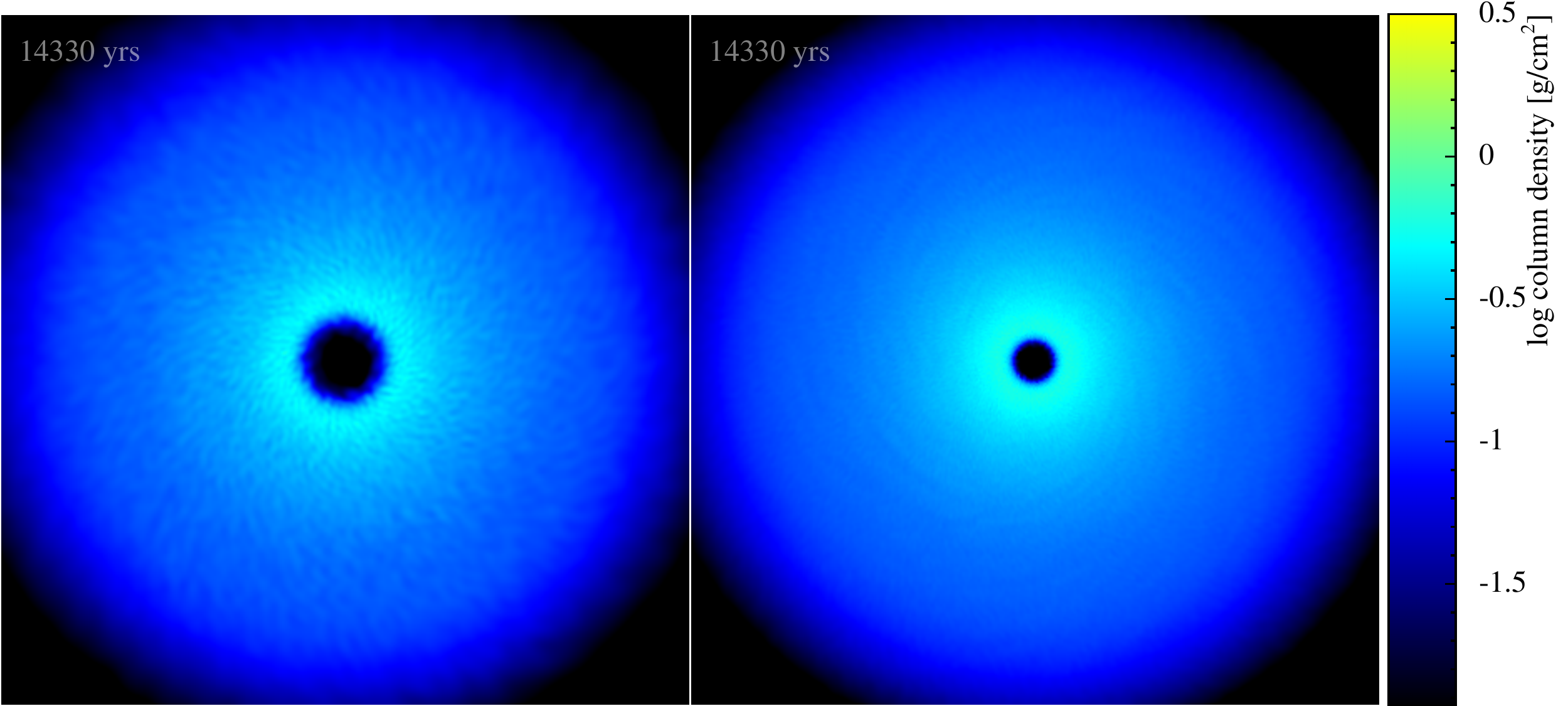}
   \caption{Rendering of the gas density of a typical T-Tauri Star protoplanetary disc at two different resolutions: $10^{5}$ (left) and $10^{6}$ (right) gas particles. Increasing the resolution smoothens the gas phase. The initial dust to gas ratio is  $0.01$ so that the dust only slightly affects the gas.}
   \label{fig:gasrender}
\end{figure*}

\begin{figure*}
  \centering
   \includegraphics[angle=0, width=\textwidth]{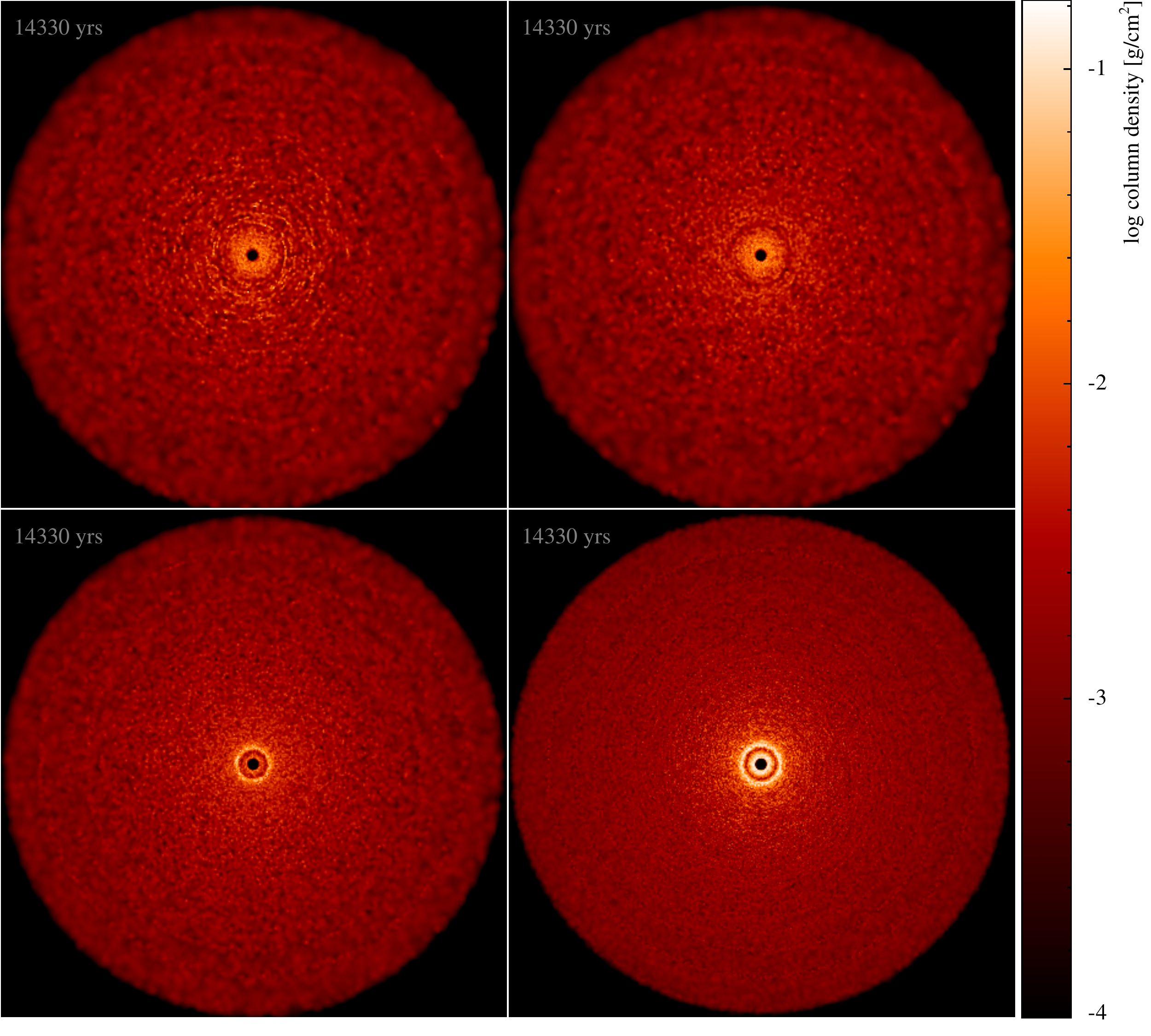}
   \caption{Rendering of the dust surface density in a typical T-Tauri Star protoplanetary disc with four different configurations: $2\times 10^{5}$ particles using a mixed smoothing length (top left) and $2\times 10^{5}$ particles (top right), $10^{5}$ dust and $10^{5}$ dust particles (bottom left) and $2\times 10^{6}$ particles (bottom right) using the gas smoothing length. Using the mixed smoothing length results in artificial structures in the dust density. Smoother dust profiles are achieved by i) using the gas smoothing length and ii) increasing the gas resolution. More accurate results are obtained by increasing the number of both the gas and the dust particles.}
   \label{fig:dustrender}
\end{figure*}

\subsection{\textsc{dustydisc}: Settling and migration in an accretion disc}
\label{sec:dustydisc}
 Our final test, \textsc{dustydisc}, concerns the evolution of the dust and gas mixture in a protoplanetary disc, where vertical settling and radial drift of the dust particles (see e.g. \citealt{Chiang2010}) are known to be crucial processes in the early stages of planet formation. SPH is well suited to this problem since i) free boundaries are trivial to implement and ii) the exact conservation of angular momentum by both the gas and dust parts of the algorithm means that the problem can be simulated for many dynamical times. We used the standard linear Epstein regime given by Eq.~\ref{eq:exprkak_Eps}. The drag force is integrated explicitly. The key features --- namely both vertical settling and radial migration --- are expected to occur. We focus here on the vertical settling of the grains since the migration is extensively discussed in \citet{Ayliffe2011}. For this specific test, the 'artificial viscosity for a disc' described in \citet{lp10} and implemented in \textsc{phantom} is used.

\subsubsection{\textsc{dustydisc}: Setup}
We setup $10^{5}$ gas particles and $10^{5}$ dust particles in a $0.01 M_{\odot}$ gas disc (with $0.0001 M_{\odot}$ of dust) surrounding a $1 M_{\odot}$ star. The disc extends from 10 to 400 AU. Both gas and dust particles are placed using a Monte-Carlo setup such that the surface density profiles of both phases are $\Sigma \left( r \right) \propto r^{-1}$. The radial profile of the gas temperature is taken to be $T\left( r \right) \propto r^{-0.6}$ with a flaring $H/r = 0.05$ at 100 AU. One code unit of time corresponds to $10^{3}$ yrs. A uniform grain size of 1 cm is used.

\subsubsection{\textsc{dustydisc}: Resolution issue}

Fig.~\ref{fig:gasrender} compares the evolution of the gas phase, varying the number of gas particles from $10^{5}$ (left panel) to $10^{6}$ (right panel). The number of dust particles does not affect the density profile of the gas given the small initial dust to gas ratio. As expected, a smoother gas profile is achieved by using a higher resolution.

Fig.~\ref{fig:dustrender} compares the evolution of the dust phase varying i) the smoothing length used to compute the drag term and ii) the number of particles in each phase. When the drag term is computed with a mixed smoothing length  ($h = [h_{\rm g} + h_{\rm d}]/2$, top left panel), artificial structures develop in the dust phase due to over-concentration of dust particles below the resolution of the gas. These numerical artefacts are removed using instead the gas smoothing length (top right panel). Indeed, the gas smoothing length is larger than the dust smoothing length since the dust grains concentrate when they reach the disc mid plane (see discussion in Sec.~\ref{sec:choiceh}). Smoother dust density profiles are achieved increasing the gas resolution ($10^{6}$ gas particles keeping $10^{5}$ dust particles, bottom left panel). Increasing the number of \textit{gas} particles thus reduces the numerical noise in the \textit{dust} phase. Finally, the smoothest dust density profile is naturally obtained when the resolution in the two phases is the highest ($2\times10^{6}$ particles, bottom right panel).

\subsubsection{\textsc{dustydisc}: Vertical settling of the particles}

\begin{figure}
  \centering
   \includegraphics[angle=0, width=\columnwidth]{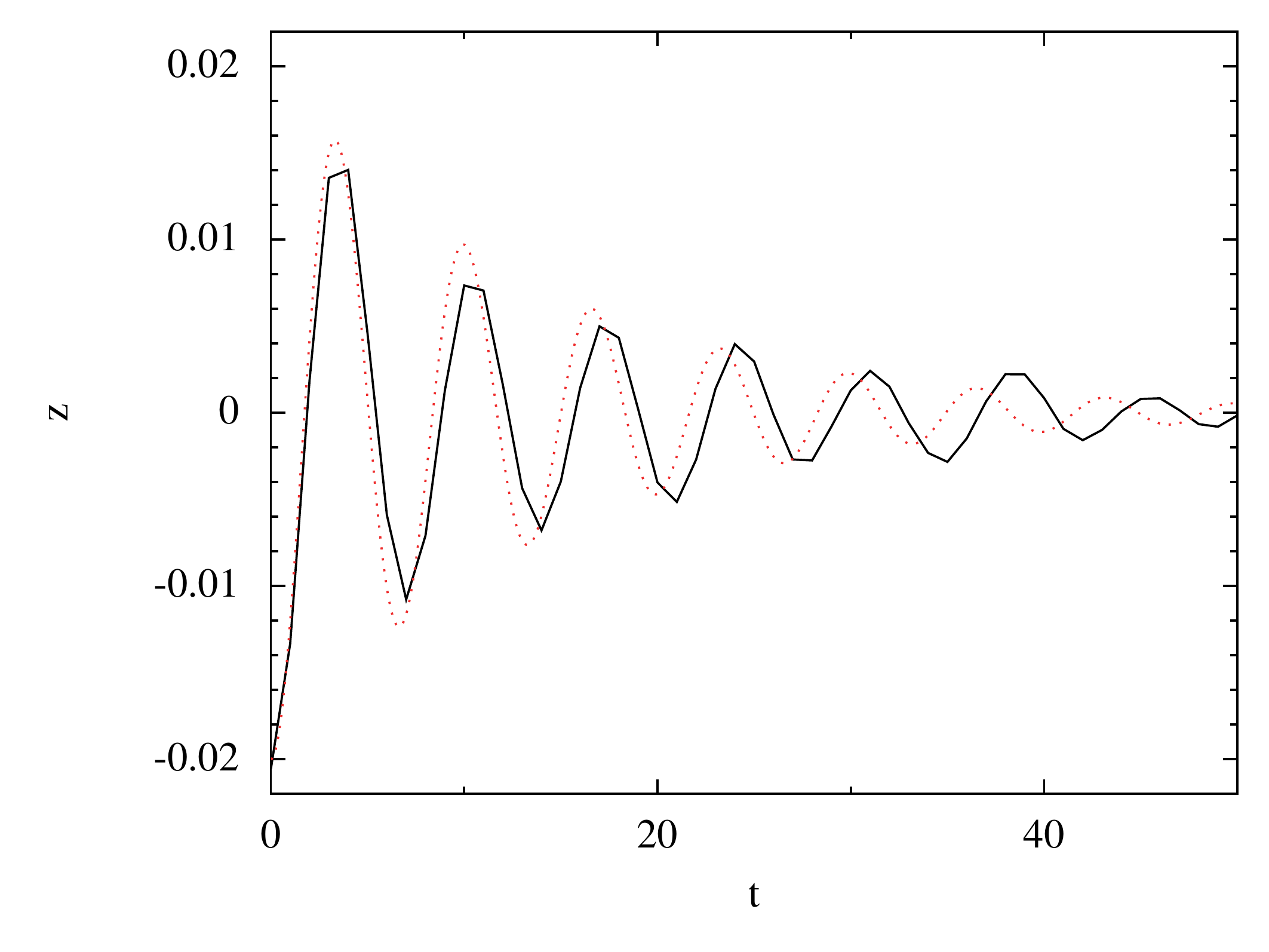}
   \caption{Vertical settling of a centimetre dust grain initially located at $r_{0}=100$ AU and $z_{0}=2$ AU (solid/black). SPH results are compared to the estimate given by the damped harmonic oscillator approximation (pointed/red). The agreement between the numerical and the analytic solutions indicates that the vertical settling of the dust grain is correctly reproduced by the SPH algorithm. The analytic estimate neglects the radial drift of the grain and the vertical stratification of the disc.}
   \label{fig:dustsettling}
\end{figure}

Fig.~\ref{fig:dustsettling} compares the vertical settling of a dust particle obtained with the SPH simulation and the analytic estimation given by the evolution of  a damped harmonic oscillator (see e.g. \citealt{GaraudLin2004}). The particle is initially located at $r=100$ AU and $z=2$ AU, and the evolution is computed for $50$ code units. The agreement between the numerical and the analytic solutions indicates that the vertical settling of the dust grain is accurately reproduced by the SPH algorithm. It should be noted that the analytic estimation neglects the radial drift of the grain and the vertical stratification of the disc. More precisely, this model assumes an expansion to order zero in $(\frac{\partial P_{\rm g}}{\partial r}/ \hrhog)/( \mathcal{G} M / r^{2})$ (meaning that the radial and the vertical motion of the dust particles are decoupled) and to first order in $z_{0}/H$ (the vertical stratification is neglected). The SPH results are thus expected to slightly differ from the analytic approximation in both amplitude and phase.

\section{Discussion}
In astrophysics, gas and dust mixtures have been predominantly studied with grid-based codes. The gas phase is computed as usual whereas the dust is treated by using superparticles (e.g. \citealt{Youdin2007}, \citealt{Bai2010}). Computing the drag is usually divided into three steps: i) interpolation of the gas velocities at the particle positions, ii) calculation of the drag force on the particles and iii) attribution of the back-reaction from the particles onto the nearby cells.

Our algorithm has two key advantages compared to the current grid-based algorithms. First, the procedure used for interpolating the gas velocities at the dust position conserves angular momentum exactly, avoiding artificial local torques. Moreover, the interpolation in current grid-based codes is performed with a standard bell-shaped kernel, regardless of the nature of the error in the drag terms. We expect that these aspects of the drag computation in grid-based codes may be improved by generalising the techniques involved in our SPH algorithm. Second, the equations of motion of the mixture (considering the drag terms only) in SPH are invariant when permuting the dust and the gas indices, i.e. $\rm{g} \leftrightarrow \rm{d}$. This symmetry is broken in grid-based schemes where the two phases are treated with two different methods (super-particles superimposed to a grid). The SPH algorithm provides rigorously identical results when interchanging the dust and the gas properties (this has been verified on the \textsc{dustybox} problem which involves only the drag, see above).

\section{Conclusions}

We have developed a new general SPH formalism for two-fluid dust and gas mixtures, with the aim of simulating the dynamics of dusty gas systems in a range of astrophysical contexts. In doing so we have generalised the standard methods developed over 15 years ago by \citet{Monaghan1995} and \citet{Monaghan1997} for treating dusty gas in SPH. In Sec.~\ref{sec:intro}, we highlighted seven key issues. In this, Paper~I, we have addressed five of these issues as follows: 1) we have introduced a simple way to compute SPH densities on two fluids with variable smoothing lengths; 2) the conservative part of the SPH equations have been derived from a Lagrangian;  3) we have demonstrated how the use of ``double-hump'' shaped kernels significantly improve the accuracy of the SPH interpolation of drag terms; 4) we find a necessary criterion $h \lesssim c_{\rm s} t_{\rm s}$ in order to correctly resolve differential motion between gas and dust that becomes critical at high drag; we also find it important to ensure $h_{\rm gas} \lesssim h_{\rm dust}$ to avoid artificial over-concentration of dust particles, implying a higher resolution should be employed in the gas phase relative to the dust.

Finally, to address issue 7), we have presented a comprehensive suite of simple test problems that can be used to benchmark astrophysical dusty gas codes. These consist of the \textsc{dustybox}, \textsc{dustywave}, \textsc{dustyshock}, \textsc{dustysedov} and \textsc{dustydisc} problems. The first three of these have known (or partially known in the case of \textsc{dustyshock}) analytic solutions and can be easily setup in any code with standard boundary conditions. We have used these tests to explore the issues raised above and have demonstrated that with the appropriate resolution criteria satisfied, our formalism is robust and provides accurate results.

The two remaining issues --- namely implicit timestepping and treatments of astrophysical drag regimes ---- are addressed in a companion paper (Paper~II). While this paper concentrates on two-fluid gas and dust mixtures, the algorithm is general and can be applied easily to the treatment of other multi-fluid systems in SPH (e.g. ambipolar diffusion).

\section*{Acknowledgments}
We thank Ben Ayliffe, Matthew Bate, Joe Monaghan and Laure Fouchet for useful discussions and comments. Figures have been produced using \textsc{splash} \citep{splashpaper} with the new \textsc{giza} backend by DJP and James Wetter. We are grateful to the Australian Research Council for funding via Discovery project grant DP1094585.

\bibliography{DustSPH,sph}

\appendix

\label{lastpage}
\end{document}

%% file: journaux.tex
%
%
%


\def\jnl@style{\it}
\def\aaref@jnl#1{{\jnl@style#1}}

\def\aaref@jnl#1{{\jnl@style#1}}

\def\aj{\aaref@jnl{AJ}}                   
\def\araa{\aaref@jnl{ARA\&A}}             
\def\apj{\aaref@jnl{ApJ}}                 
\def\apjl{\aaref@jnl{ApJ}}                
\def\apjs{\aaref@jnl{ApJS}}               
\def\ao{\aaref@jnl{Appl.~Opt.}}           
\def\apss{\aaref@jnl{Ap\&SS}}             
\def\aap{\aaref@jnl{A\&A}}                
\def\aapr{\aaref@jnl{A\&A~Rev.}}          
\def\aaps{\aaref@jnl{A\&AS}}              
\def\azh{\aaref@jnl{AZh}}                 
\def\baas{\aaref@jnl{BAAS}}               
\def\jrasc{\aaref@jnl{JRASC}}             
\def\memras{\aaref@jnl{MmRAS}}            
\def\mnras{\aaref@jnl{MNRAS}}             
\def\pra{\aaref@jnl{Phys.~Rev.~A}}        
\def\prb{\aaref@jnl{Phys.~Rev.~B}}        
\def\prc{\aaref@jnl{Phys.~Rev.~C}}        
\def\prd{\aaref@jnl{Phys.~Rev.~D}}        
\def\pre{\aaref@jnl{Phys.~Rev.~E}}        
\def\prl{\aaref@jnl{Phys.~Rev.~Lett.}}    
\def\pasp{\aaref@jnl{PASP}}               
\def\pasj{\aaref@jnl{PASJ}}               
\def\qjras{\aaref@jnl{QJRAS}}             
\def\skytel{\aaref@jnl{S\&T}}             
\def\solphys{\aaref@jnl{Sol.~Phys.}}      
\def\sovast{\aaref@jnl{Soviet~Ast.}}      
\def\ssr{\aaref@jnl{Space~Sci.~Rev.}}     
\def\zap{\aaref@jnl{ZAp}}                 
\def\nat{\aaref@jnl{Nature}}              
\def\iaucirc{\aaref@jnl{IAU~Circ.}}       
\def\aplett{\aaref@jnl{Astrophys.~Lett.}} 
\def\apspr{\aaref@jnl{Astrophys.~Space~Phys.~Res.}}
\def\bain{\aaref@jnl{Bull.~Astron.~Inst.~Netherlands}} 
\def\fcp{\aaref@jnl{Fund.~Cosmic~Phys.}}  
\def\gca{\aaref@jnl{Geochim.~Cosmochim.~Acta}}   
\def\grl{\aaref@jnl{Geophys.~Res.~Lett.}} 
\def\jcp{\aaref@jnl{J.~Chem.~Phys.}}      
\def\jgr{\aaref@jnl{J.~Geophys.~Res.}}    
\def\jqsrt{\aaref@jnl{J.~Quant.~Spec.~Radiat.~Transf.}}
\def\memsai{\aaref@jnl{Mem.~Soc.~Astron.~Italiana}}
\def\nphysa{\aaref@jnl{Nucl.~Phys.~A}}   
\def\physrep{\aaref@jnl{Phys.~Rep.}}   
\def\physscr{\aaref@jnl{Phys.~Scr}}   
\def\planss{\aaref@jnl{Planet.~Space~Sci.}}   
\def\procspie{\aaref@jnl{Proc.~SPIE}}   

\let\astap=\aap
\let\apjlett=\apjl
\let\apjsupp=\apjs
\let\applopt=\ao

%% file: Dust_I.bbl
\begin{thebibliography}{}

\bibitem[\protect\citeauthoryear{{Ayliffe}, {Laibe}, {Price} \&
  {Bate}}{{Ayliffe} et~al.}{2011}]{Ayliffe2011}
{Ayliffe} B.,  {Laibe} G.,  {Price} D.~J.,    {Bate} M.~R.,  2011, MNRAS,
  submitted

\bibitem[\protect\citeauthoryear{{Bai} \& {Stone}}{{Bai} \&
  {Stone}}{2010}]{Bai2010}
{Bai} X.-N.,  {Stone} J.~M.,  2010, \apjs, 190, 297

\bibitem[\protect\citeauthoryear{{Baines}, {Williams} \& {Asebiomo}}{{Baines}
  et~al.}{1965}]{Baines1965}
{Baines} M.~J.,  {Williams} I.~P.,    {Asebiomo} A.~S.,  1965, \mnras, 130, 63

\bibitem[\protect\citeauthoryear{{Barri{\`e}re-Fouchet}, {Gonzalez}, {Murray},
  {Humble} \& {Maddison}}{{Barri{\`e}re-Fouchet} et~al.}{2005}]{BF2005}
{Barri{\`e}re-Fouchet} L.,  {Gonzalez} J.-F.,  {Murray} J.~R.,  {Humble} R.~J.,
     {Maddison} S.~T.,  2005, \aap, 443, 185

\bibitem[\protect\citeauthoryear{{Chiang} \& {Youdin}}{{Chiang} \&
  {Youdin}}{2010a}]{ChiangYoudin2010}
{Chiang} E.,  {Youdin} A.~N.,  2010a, Annual Review of Earth and Planetary
  Sciences, 38, 493

\bibitem[\protect\citeauthoryear{{Chiang} \& {Youdin}}{{Chiang} \&
  {Youdin}}{2010b}]{Chiang2010}
{Chiang} E.,  {Youdin} A.~N.,  2010b, Annual Review of Earth and Planetary
  Sciences, 38, 493

\bibitem[\protect\citeauthoryear{{Epstein}}{{Epstein}}{1924}]{Epstein1924}
{Epstein} P.~S.,  1924, Physical Review, 23, 710

\bibitem[\protect\citeauthoryear{{Fan} \& {Zhu}}{{Fan} \& {Zhu}}{1998}]{FanZhu}
{Fan} L.-S.,  {Zhu} C.,  1998, Principles of Gas-Solid Flows.
Cambridge University Press

\bibitem[\protect\citeauthoryear{{Fromang} \& {Nelson}}{{Fromang} \&
  {Nelson}}{2005}]{Fromang2005}
{Fromang} S.,  {Nelson} R.~P.,  2005, \mnras, 364, L81

\bibitem[\protect\citeauthoryear{{Fromang} \& {Papaloizou}}{{Fromang} \&
  {Papaloizou}}{2006}]{Fromang2006}
{Fromang} S.,  {Papaloizou} J.,  2006, \aap, 452, 751

\bibitem[\protect\citeauthoryear{{Fulk} \& {Quinn}}{{Fulk} \&
  {Quinn}}{1996}]{fq96}
{Fulk} D.~A.,  {Quinn} D.~W.,  1996, J. Comp. Phys., 126, 165

\bibitem[\protect\citeauthoryear{{Garaud} \& {Lin}}{{Garaud} \&
  {Lin}}{2004}]{GaraudLin2004}
{Garaud} P.,  {Lin} D.~N.~C.,  2004, \apj, 608, 1050

\bibitem[\protect\citeauthoryear{{Goodman} \& {Pindor}}{{Goodman} \&
  {Pindor}}{2000}]{Goodman2000}
{Goodman} J.,  {Pindor} B.,  2000, Icarus, 148, 537

\bibitem[\protect\citeauthoryear{{Harlow} \& {Amsden}}{{Harlow} \&
  {Amsden}}{1975}]{harlow1975}
{Harlow} F.~H.,  {Amsden} A.~A.,  1975, J. Comp. Phys., 17, 19

\bibitem[\protect\citeauthoryear{{Iannuzzi} \& {Dolag}}{{Iannuzzi} \&
  {Dolag}}{2011}]{id2011}
{Iannuzzi} F.,  {Dolag} K.,  2011, \mnras, submitted preprint

\bibitem[\protect\citeauthoryear{{Johansen}, {Oishi}, {Low}, {Klahr}, {Henning}
  \& {Youdin}}{{Johansen} et~al.}{2007}]{Johansen2007}
{Johansen} A.,  {Oishi} J.~S.,  {Low} M.,  {Klahr} H.,  {Henning} T.,
  {Youdin} A.,  2007, \nat, 448, 1022

\bibitem[\protect\citeauthoryear{{Laibe} \& {Price}}{{Laibe} \&
  {Price}}{2011c}]{LP11c}
{Laibe} G.,  {Price} D.~J.,  2011c, MNRAS, in press (Paper II)

\bibitem[\protect\citeauthoryear{{Laibe} \& {Price}}{{Laibe} \&
  {Price}}{2011a}]{LP11a}
{Laibe} G.,  {Price} D.~J.,  2011a, MNRAS, in press

\bibitem[\protect\citeauthoryear{{Lodato} \& {Price}}{{Lodato} \&
  {Price}}{2010}]{lp10}
{Lodato} G.,  {Price} D.~J.,  2010, MNRAS, 405, 1212

\bibitem[\protect\citeauthoryear{{Maddison}, {Humble} \& {Murray}}{{Maddison}
  et~al.}{2003}]{Maddison2003}
{Maddison} S.~T.,  {Humble} R.~J.,    {Murray} J.~R.,  2003, in {D.~Deming \&
  S.~Seager} ed., Scientific Frontiers in Research on Extrasolar Planets
  Vol.~294 of Astronomical Society of the Pacific Conference Series, {Building
  Planets with Dusty Gas}.
pp 307--310

\bibitem[\protect\citeauthoryear{{Marble}}{{Marble}}{1970}]{Marble1970}
{Marble} F.~E.,  1970, Ann. Rev. Fluid Mech., 2, 397

\bibitem[\protect\citeauthoryear{{Merlin}, {Buonomo}, {Grassi}, {Piovan} \&
  {Chiosi}}{{Merlin} et~al.}{2010}]{Merlin2010}
{Merlin} E.,  {Buonomo} U.,  {Grassi} T.,  {Piovan} L.,    {Chiosi} C.,  2010,
  \aap, 513, A36

\bibitem[\protect\citeauthoryear{{Miniati}}{{Miniati}}{2010}]{Miniati2010}
{Miniati} F.,  2010, Journal of Computational Physics, 229, 3916

\bibitem[\protect\citeauthoryear{{Miura} \& {Glass}}{{Miura} \&
  {Glass}}{1982}]{Miura1982}
{Miura} H.,  {Glass} I.~I.,  1982, Roy. Soc. Lon. Proc. Ser. A, 382, 373

\bibitem[\protect\citeauthoryear{{Monaghan}}{{Monaghan}}{1992}]{monaghan92}
{Monaghan} J.~J.,  1992, Ann. Rev. Astron. Astrophys., 30, 543

\bibitem[\protect\citeauthoryear{{Monaghan}}{{Monaghan}}{1997}]{Monaghan1997}
{Monaghan} J.~J.,  1997, Journal of Computational Physics, 138, 801

\bibitem[\protect\citeauthoryear{{Monaghan} \& {Kocharyan}}{{Monaghan} \&
  {Kocharyan}}{1995}]{Monaghan1995}
{Monaghan} J.~J.,  {Kocharyan} A.,  1995, Computer Physics Communications, 87,
  225

\bibitem[\protect\citeauthoryear{{Morris}}{{Morris}}{1996a}]{morris96}
{Morris} J.~P.,  1996a, PASA, 13, 97

\bibitem[\protect\citeauthoryear{{Morris}}{{Morris}}{1996b}]{morrisphd}
{Morris} J.~P.,  1996b, PhD thesis, {Monash University, Melbourne, Australia}

\bibitem[\protect\citeauthoryear{{Paardekooper} \& {Mellema}}{{Paardekooper} \&
  {Mellema}}{2006}]{PM2006}
{Paardekooper} S.-J.,  {Mellema} G.,  2006, \aap, 453, 1129

\bibitem[\protect\citeauthoryear{{Price}}{{Price}}{2007}]{splashpaper}
{Price} D.~J.,  2007, Publ. Astron. Soc. Aust., 24, 159

\bibitem[\protect\citeauthoryear{{Price}}{{Price}}{2011}]{Price2011}
{Price} D.~J.,  2011, J. Comp. Phys., in press, doi:10.1016/j.jcp.2010.12.011

\bibitem[\protect\citeauthoryear{{Price} \& {Federrath}}{{Price} \&
  {Federrath}}{2010}]{pf10}
{Price} D.~J.,  {Federrath} C.,  2010, MNRAS, 406, 1659

\bibitem[\protect\citeauthoryear{{Price} \& {Monaghan}}{{Price} \&
  {Monaghan}}{2004}]{pm04b}
{Price} D.~J.,  {Monaghan} J.~J.,  2004, MNRAS, 348, 139

\bibitem[\protect\citeauthoryear{{Price} \& {Monaghan}}{{Price} \&
  {Monaghan}}{2005}]{pm05}
{Price} D.~J.,  {Monaghan} J.~J.,  2005, MNRAS, 364, 384

\bibitem[\protect\citeauthoryear{{Price} \& {Monaghan}}{{Price} \&
  {Monaghan}}{2007}]{pm07}
{Price} D.~J.,  {Monaghan} J.~J.,  2007, MNRAS, 374, 1347

\bibitem[\protect\citeauthoryear{{Rice}, {Lodato}, {Pringle}, {Armitage} \&
  {Bonnell}}{{Rice} et~al.}{2004}]{Rice04}
{Rice} W.~K.~M.,  {Lodato} G.,  {Pringle} J.~E.,  {Armitage} P.~J.,
  {Bonnell} I.~A.,  2004, \mnras, 355, 543

\bibitem[\protect\citeauthoryear{{Rosswog} \& {Price}}{{Rosswog} \&
  {Price}}{2007}]{rp07}
{Rosswog} S.,  {Price} D.,  2007, MNRAS, 379, 915

\bibitem[\protect\citeauthoryear{Rudinger}{Rudinger}{1964}]{Rudinger1964}
Rudinger 1964, Physics of Fluids, 7, 658

\bibitem[\protect\citeauthoryear{Saito, Marumoto \& Takayama}{Saito
  et~al.}{2003}]{Saito2003}
Saito T.,  Marumoto M.,    Takayama K.,  2003, Shock Waves, 13, 299

\bibitem[\protect\citeauthoryear{{Sod}}{{Sod}}{1978}]{sod78}
{Sod} G.~A.,  1978, J. Comp. Phys., 27, 1

\bibitem[\protect\citeauthoryear{{Springel} \& {Hernquist}}{{Springel} \&
  {Hernquist}}{2002}]{sh02}
{Springel} V.,  {Hernquist} L.,  2002, MNRAS, 333, 649

\bibitem[\protect\citeauthoryear{{Youdin} \& {Johansen}}{{Youdin} \&
  {Johansen}}{2007}]{Youdin2007}
{Youdin} A.,  {Johansen} A.,  2007, \apj, 662, 613

\bibitem[\protect\citeauthoryear{{Youdin} \& {Goodman}}{{Youdin} \&
  {Goodman}}{2005}]{Youdin2005}
{Youdin} A.~N.,  {Goodman} J.,  2005, \apj, 620, 459

\end{thebibliography}
